\documentclass[amsmath,amssymb,11pt,tightenlines,superscriptaddress,nofootinbib,preprintnumbers, notitlepage,aps,prd,]{revtex4-1}

\usepackage{multirow}
\usepackage{amsmath,amssymb}
\usepackage{graphicx}
\usepackage{caption}
\usepackage{subcaption}
\usepackage{feynmp}
\usepackage{color}
\usepackage{url}
\usepackage{slashed}
\usepackage{footmisc}
\usepackage{hyperref}
\usepackage{ragged2e}
\DeclareCaptionJustification{justified}{\justifying}
\begin{document}
% hyper-link hep-ph references
\newcommand{\arxiv}[1]{\href{http://arxiv.org/abs/#1}{arXiv:#1}}

\def\contentsname{{\normalsize Content}}
\def\tablename{Table}
\def\figurename{Figure}

\def\bdrs{\text{BDRS}}
\def\pveto{P_\text{veto}}
\def\nj{n_\text{jets}}
\def\meff{m_\text{eff}}
\def\ptmin{p_T^\text{min}}
\def\gtot{\Gamma_\text{tot}}
\def\as{\alpha_s}
\def\az{\alpha_0}
\def\gz{g_0}
\def\w{\vec{w}}
\def\sdag{\Sigma^{\dag}}
\def\s{\Sigma}
\newcommand{\psib}{\overline{\psi}}
\newcommand{\Psib}{\overline{\Psi}}
\newcommand\one{\leavevmode\hbox{\small1\normalsize\kern-.33em1}}
\newcommand{\Mpl}{M_\mathrm{Pl}}
\newcommand{\p}{\partial}
\newcommand{\mat}{\mathcal{M}}
\newcommand{\lag}{\mathcal{L}}
\newcommand{\ord}{\mathcal{O}}
\newcommand{\ope}{\mathcal{O}}
\newcommand{\qqquad}{\qquad \qquad}
\newcommand{\qqqquad}{\qquad \qquad \qquad}

\newcommand{\qb}{\bar{q}}
\newcommand{\matx}{|\mathcal{M}|^2}
\newcommand{\really}{\stackrel{!}{=}}
\newcommand{\msbar}{\overline{\text{MS}}}
\newcommand{\qns}{f_q^\text{NS}}
\newcommand{\lqcd}{\Lambda_\text{QCD}}
\newcommand{\met}{\slashchar{E}_T}
\newcommand{\pmiss}{\slashchar{\vec{p}}_T}

\newcommand{\sq}{\tilde{q}}
\newcommand{\go}{\tilde{g}}
\newcommand{\st}[1]{\tilde{t}_{#1}}
\newcommand{\stb}[1]{\tilde{t}_{#1}^*}
\newcommand{\nz}[1]{\tilde{\chi}_{#1}^0}
\newcommand{\cp}[1]{\tilde{\chi}_{#1}^+}
\newcommand{\CP}{CP}

% all the masses 
\providecommand{\mg}{m_{\tilde{g}}}
\providecommand{\mst}[1]{m_{\tilde{t}_{#1}}}
\newcommand{\msn}[1]{m_{\tilde{\nu}_{#1}}}
\newcommand{\mch}[1]{m_{\tilde{\chi}^+_{#1}}}
\newcommand{\mne}[1]{m_{\tilde{\chi}^0_{#1}}}
\newcommand{\msb}[1]{m_{\tilde{b}_{#1}}}
\newcommand{\vsm}{\ensuremath{v_{\rm SM}}}

% units of measure
\newcommand{\mev}{{\ensuremath\rm MeV}}
\newcommand{\gev}{{\ensuremath\rm GeV}}
\newcommand{\tev}{{\ensuremath\rm TeV}}
\newcommand{\sign}{{\ensuremath\rm sign}}
\newcommand{\iab}{{\ensuremath\rm ab^{-1}}}
\newcommand{\ifb}{{\ensuremath\rm fb^{-1}}}
\newcommand{\ipb}{{\ensuremath\rm pb^{-1}}}

% really great macro by Chris Lester
\def\slashchar#1{\setbox0=\hbox{$#1$}           % set a box for #1
   \dimen0=\wd0                                 % and get its size
   \setbox1=\hbox{/} \dimen1=\wd1               % get size of /
   \ifdim\dimen0>\dimen1                        % #1 is bigger
      \rlap{\hbox to \dimen0{\hfil/\hfil}}      % so center / in box
      #1                                        % and print #1
   \else                                        % / is bigger
      \rlap{\hbox to \dimen1{\hfil$#1$\hfil}}   % so center #1
      /                                         % and print /
   \fi}
\newcommand{\dslash}{\slashchar{\partial}}
\newcommand{\Dslash}{\slashchar{D}}

\newcommand{\eg}{\textsl{e.g.}\;}
\newcommand{\ie}{\textsl{i.e.}\;}
\newcommand{\etal}{\textsl{et al}\;}
%\DeclareMathOperator{\tr}{Tr}

% maximal number of floating environments on each page 
\setlength{\floatsep}{0pt}
\setcounter{topnumber}{1}
\setcounter{bottomnumber}{1}
\setcounter{totalnumber}{1}
\renewcommand{\topfraction}{1.0}
\renewcommand{\bottomfraction}{1.0}
\renewcommand{\textfraction}{0.0}

\newcommand{\rig}{\rightarrow}
\newcommand{\lrig}{\longrightarrow}
\renewcommand{\d}{{\mathrm{d}}}
\newcommand{\be}{\begin{eqnarray*}}
\newcommand{\ee}{\end{eqnarray*}}
\newcommand{\gl}[1]{(\ref{#1})}
\newcommand{\ta}[2]{ \frac{ {\mathrm{d}} #1 } {{\mathrm{d}} #2}}
\newcommand{\bee}{\begin{eqnarray}}
\newcommand{\eee}{\end{eqnarray}}
\newcommand{\beeq}{\begin{equation}}
\newcommand{\eeeq}{\end{equation}}
\newcommand{\mc}{\mathcal}
\newcommand{\mr}{\mathrm}
\newcommand{\ep}{\varepsilon}
\newcommand{\emt}{$\times 10^{-3}$}
\newcommand{\emfo}{$\times 10^{-4}$}
\newcommand{\emfi}{$\times 10^{-5}$}

\newcommand{\revision}[1]{{\bf{}#1}}

\newcommand{\hzero}{h^0}
\newcommand{\Hzero}{H^0}
\newcommand{\Azero}{A^0}
\newcommand{\PHiggs}{H}
\newcommand{\PW}{W}
\newcommand{\PZ}{Z}

\newcommand{\sw}{\ensuremath{s_w}}
\newcommand{\cw}{\ensuremath{c_w}}
\newcommand{\swd}{\ensuremath{s^2_w}}
\newcommand{\cwd}{\ensuremath{c^2_w}}

%% 2HDM Higgs masses
\newcommand{\mhhd}{\ensuremath{m^2_{\Hzero}}}
\newcommand{\mhh}{\ensuremath{m_{\Hzero}}}
\newcommand{\mlhd}{\ensuremath{m^2_{\hzero}}}
\newcommand{\Mlh}{\ensuremath{m_{\hzero}}}
\newcommand{\mad}{\ensuremath{m^2_{\Azero}}}
\newcommand{\mhpd}{\ensuremath{m^2_{\PHiggs^{\pm}}}}
\newcommand{\mhp}{\ensuremath{m_{\PHiggs^{\pm}}}}

 \newcommand{\sa}{\ensuremath{\sin\alpha}}
 \newcommand{\ca}{\ensuremath{\cos\alpha}}
 \newcommand{\cad}{\ensuremath{\cos^2\alpha}}
 \newcommand{\sad}{\ensuremath{\sin^2\alpha}}
 \newcommand{\sbd}{\ensuremath{\sin^2\beta}}
 \newcommand{\cbd}{\ensuremath{\cos^2\beta}}
 \newcommand{\cb}{\ensuremath{\cos\beta}}
 \renewcommand{\sb}{\ensuremath{\sin\beta}}
 \newcommand{\tanbd}{\ensuremath{\tan^2\beta}}
 \newcommand{\cotbd}{\ensuremath{\cot^2\beta}}
 \newcommand{\tanb}{\ensuremath{\tan\beta}}
 \newcommand{\tb}{\ensuremath{\tan\beta}}
 \newcommand{\cotb}{\ensuremath{\cot\beta}}

%\newcommand{\GeV}{\ensuremath{\rm GeV}}
%\newcommand{\MeV}{\ensuremath{\rm MeV}}
%\newcommand{\TeV}{\ensuremath{\rm TeV}}

%%% Local Variables:
%%% mode: latex
%%% TeX-master: "paper"
%%% End:

\begin{fmffile}{feyn}

\preprint{PITT-PACC-1702}
\title{Weak Boson Fusion at 100~TeV}

\author{Dorival Gon\c{c}alves} 
\affiliation{PITT-PACC,  Department  of  Physics  and Astronomy,  University  of  Pittsburgh,  USA}
\author{Tilman Plehn}
\affiliation{Institut f\"ur Theoretische Physik, Universit\"at Heidelberg, Germany}
\author{Jennifer M. Thompson}
\affiliation{Institut f\"ur Theoretische Physik, Universit\"at Heidelberg, Germany}

\begin{abstract}
  From the LHC runs we know that, with increasing collider energy,
  weak-boson-fusion Higgs production dominates as an environment for
  precision measurements. We show how a future hadron collider
  performs for three challenging benchmark signatures. Because all of
  these measurements rely on the tagging jet signature, we first give
  a comprehensive analysis of weak-boson-fusion kinematics and a
  proposed two-step jet veto at a 100 TeV hadron collider.  We then
  find this machine to be sensitive to invisible Higgs branching
  ratios of 0.5\%, a second-generation muon Yukawa coupling of 2\%,
  and an enhanced total Higgs width of around 5\%, the latter with
  essentially no model dependence. This kind of performance crucially
  relies on a sufficient detector coverage and a dedicated
  weak-boson-fusion trigger channel.
\end{abstract}

\maketitle
\tableofcontents

\clearpage
%%%%%%%%%%%%%%%%%%%%%%%%%%%%%%%%%%%%%%%%%%%%%%%%%%
\section{Introduction}
\label{sec:intro}

After the discovery of a light and likely fundamental Higgs
boson~\cite{higgs,discovery}, one of the main goals of particle
physics is to test how well the Standard Model describes this particle
and its different properties~\cite{higgs_fit}. Beyond the LHC time
scale, searches for new physics in the newly discovered Higgs sector
are one of the main driving forces behind new colliders. While the
expected precision of an $e^+ e^-$ Higgs factory has been studied in
some detail~\cite{sfitter_future}, the corresponding results for a
future hadron collider~\cite{FCC-hh,SppC} is not yet available. One
reason for this is that we expect precision measurements in
essentially all standard Higgs channels to be limited by experimental
systematics and theoretical uncertainties. A global analysis would
simply translate guesses on these two inputs into a highly speculative
estimate of the physics reach. On the other hand, we can identify a
set of benchmark channels which are not entirely theory or
systematics limited. For some of these channels we will illustrate the
power of a future hadron collider in Higgs precision studies in this
paper.\medskip

The physics goals and opportunities of a 100~TeV hadron
collider~\cite{FCC-hh,SppC,fcc_higgs,fcc_bsm} with an integrated
luminosity around $20~\iab$~\cite{FCClumi} are currently under intense
investigation. A leading pillar of its physics program will be studies
of weakly interacting thermal dark matter~\cite{DM100TeV}; it will, for
example, be complemented by searches for heavy Higgs
bosons~\cite{BSMhiggs}, studies of the electroweak gauge sector at
high energies~\cite{WZ100TeV}, and tests of the nature of the
electroweak phase transition~\cite{EWPT}. In the Higgs sector two
crucial measurements, which can be reliably studied, are of the top
Yukawa coupling~\cite{Top100TeV} and of the triple Higgs
coupling~\cite{Self100TeV,hh_wbf}.\medskip

We will ask three additional questions, all of which are related to weak
boson fusion production of a SM-like Higgs~\cite{dave_thesis}.  This
production process is known to be highly efficient at the LHC when
combined with standard Higgs decays to tau leptons~\cite{wbf_tau} or
to $W$-bosons~\cite{wbf_w}. Its theoretical description is more
precise than almost any other process at the LHC~\cite{wbf_loops}. The
only reason why it played hardly any role in the Higgs discovery was
the reduced LHC energy during Run~I. Moreover, with these signatures
it has been, from the very beginning, at the heart of Higgs precision
analyses~\cite{dieter_duehrssen}. A critical ingredient to the success
of weak boson fusion as a Higgs production channel is the central jet
veto~\cite{cjv}. This removes a large proportion of the QCD backgrounds,
which would otherwise overwhelm any analysis. We will discuss it in detail in
Sec.~\ref{sec:jets}.

First, an invisible Higgs decay to a pair of dark matter particles is
not only an obvious channel to search for, it is also very well
motivated for example in Higgs portal models~\cite{inv_portal} and in
supersymmetric extensions of the Standard Model~\cite{inv_nmssm}. At
the LHC we will be able to probe invisible Higgs branching ratios in
the few per-cent range~\cite{eboli_zeppenfeld,cathy}, and in
Sec.~\ref{sec:inv} we will see how much better a future hadron
collider will be able to do in this decay channel.

Second, the LHC will firmly establish that the Higgs couples to the
fermions of the third generation, but the size of the Higgs couplings
to the second generation fermions will remain largely unknown. While
there are new ideas to measure the Yukawa couplings to
second-generation quarks~\cite{second_quark}, the obvious task is to
measure the Higgs branching ratio to muons~\cite{lhc_muons}. We will
show in Sec.~\ref{sec:muons} how a 100~TeV collider will turn a proof
of the existence of such a coupling into a precise measurement.

Finally, one of the main drawbacks of a hadron collider has always
been that it does not allow for a direct measurement of the Higgs
width. This changes when we include off-shell Higgs production, for
example in the four-lepton final state~\cite{offshell}. The problem
with this measurement in gluon fusion is that it relies on the
assumption that the effective Higgs-gluon coupling has an energy
scaling like in the Standard Model~\cite{offshell_model}. At a 100~TeV
collider we can instead use weak-boson-fusion production with the
known, logarithmic scaling given by the renormalization group running
of the weak coupling, as we will discuss in Sec.~\ref{sec:offshell}. A
comprehensive analysis of Higgs pair production in weak boson fusion
will not be part of our analysis, but can be found in
Ref.~\cite{hh_wbf}.

%%%%%%%%%%%%%%%%%%%%%%%%%%%%%%%%%%%%%%%%%%%%%%%%%%
\section{Tagging jets and jet veto}
\label{sec:jets}

 %-------------------------------------------------------
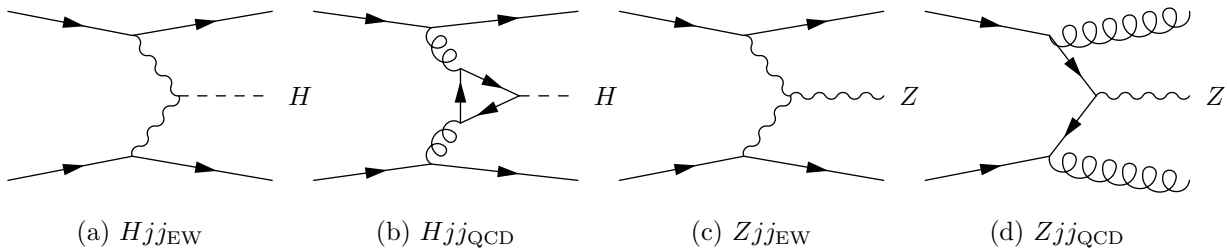
\begin{figure}[b!]
\begin{subfigure}[b]{0.24\textwidth}
 %Hjj EW
\begin{fmfgraph*}(100,80)
\fmfset{arrow_len}{3mm}
\fmftop{q1,q3}
\fmfbottom{q2,q4}
\fmfright{H}
\fmflabel{{$H$}}{H}
\fmf{fermion,tension=1.5,width=0.5}{q1,v1}
\fmf{fermion,width=0.5}{v1,q3}
\fmf{photon,width=0.5}{v1,v3,v4}
\fmf{dashes,width=0.5}{v3,H}
\fmf{fermion,tension=1.5,width=0.5}{q2,v4}
\fmf{fermion,width=0.5}{v4,q4}
%\fmfv{decor.shape=circle,decor.filled=full,decor.size=2.5}{v3}
\end{fmfgraph*}
\caption{$Hjj_\text{EW}$}
\end{subfigure}
%Hjj QCD
\begin{subfigure}[b]{0.24\textwidth}
\begin{fmfgraph*}(100,80)
\fmfset{arrow_len}{3mm}
\fmftop{q1,q3}
\fmfbottom{q2,q4}
\fmfright{H}
\fmflabel{{$H$}}{H}
\fmf{fermion,tension=1.5,width=0.5}{q1,v1}
\fmf{fermion,width=0.5}{v1,q3}
\fmf{gluon,width=0.5}{v1,v5}
\fmf{fermion,width=0.5,tension=0.5}{v5,v3,v6,v5}
\fmf{gluon,width=0.5}{v6,v4}
\fmf{dashes,width=0.5}{v3,H}
\fmf{fermion,tension=1.5,width=0.5}{q2,v4}
\fmf{fermion,width=0.5}{v4,q4}
%\fmfv{decor.shape=circle,decor.filled=full,decor.size=2.5}{v3}
\end{fmfgraph*}
\caption{$Hjj_\text{QCD}$}
\end{subfigure}
 %Zjj EW
\begin{subfigure}[b]{0.24\textwidth}
\begin{fmfgraph*}(100,80)
\fmfset{arrow_len}{3mm}
\fmftop{q1,q3}
\fmfbottom{q2,q4}
\fmfright{Z}
\fmflabel{{$Z$}}{Z}
\fmf{fermion,tension=1.5,width=0.5}{q1,v1}
\fmf{fermion,width=0.5}{v1,q3}
\fmf{photon,width=0.5}{v1,v3,v4}
\fmf{photon,width=0.5}{v3,Z}
\fmf{fermion,tension=1.5,width=0.5}{q2,v4}
\fmf{fermion,width=0.5}{v4,q4}
%\fmfv{decor.shape=circle,decor.filled=full,decor.size=2.5}{v3}
\end{fmfgraph*}
\caption{$Zjj_\text{EW}$}
\end{subfigure}
%Zjj QCD
\begin{subfigure}[b]{0.24\textwidth}
\begin{fmfgraph*}(100,80)
\fmfset{arrow_len}{3mm}
\fmftop{q1,q3}
\fmfbottom{q2,q4}
\fmfright{Z}
\fmflabel{{$Z$}}{Z}
\fmf{fermion,tension=1.5,width=0.5}{q1,v1}
\fmf{gluon,width=0.5}{v1,q3}
\fmf{fermion,width=0.5}{v1,v3,v4}
\fmf{photon,width=0.5}{v3,Z}
\fmf{fermion,tension=1.5,width=0.5}{q2,v4}
\fmf{gluon,width=0.5}{v4,q4}
%\fmfv{decor.shape=circle,decor.filled=full,decor.size=2.5}{v3}
\end{fmfgraph*}
\caption{$Zjj_\text{QCD}$}
    \end{subfigure}
 \caption{Representative set of Feynman diagrams for the WBF Higgs
   signal, the contribution from gluon fusion Higgs production, and
   the two $Z$+jets backgrounds.}
 \label{fig:feyn}
\end{figure}
%-------------------------------------------------------

A proper understanding and use of the two tagging jets is crucial to
any weak boson fusion (WBF) analysis at the LHC or at a future hadron
collider. The signal(s) and at least some irreducible backgrounds are
independent of the Higgs decay channel, as shown in
Fig.~\ref{fig:feyn}. Unless otherwise noted, we generate all signal
and background events with \textsc{Sherpa}~\cite{sherpa}, merged up to
three jets through the \textsc{CKKW} algorithm~\cite{ckkw}, and
accounting for hadronization effects. The two tagging jet candidates 
are defined as the two hardest anti-$k_T$ jets with $R =
0.4$ and $p_{T,j}>40$~GeV, obtained with
\textsc{Fastjet}~\cite{fastjet}.  The full top-mass dependence for
gluon-fusion Higgs production is included through re-weighting the
effective field theory to the full calculation at each phase-space
point. The loop contributions are provided by
\textsc{OpenLoops}~\cite{openloops}. The scales are set according to
the \textsc{Sherpa Mets} scale setting
algorithm~\cite{sherpa}.\medskip

\begin{figure}[t!]
\includegraphics[width=.35\textwidth]{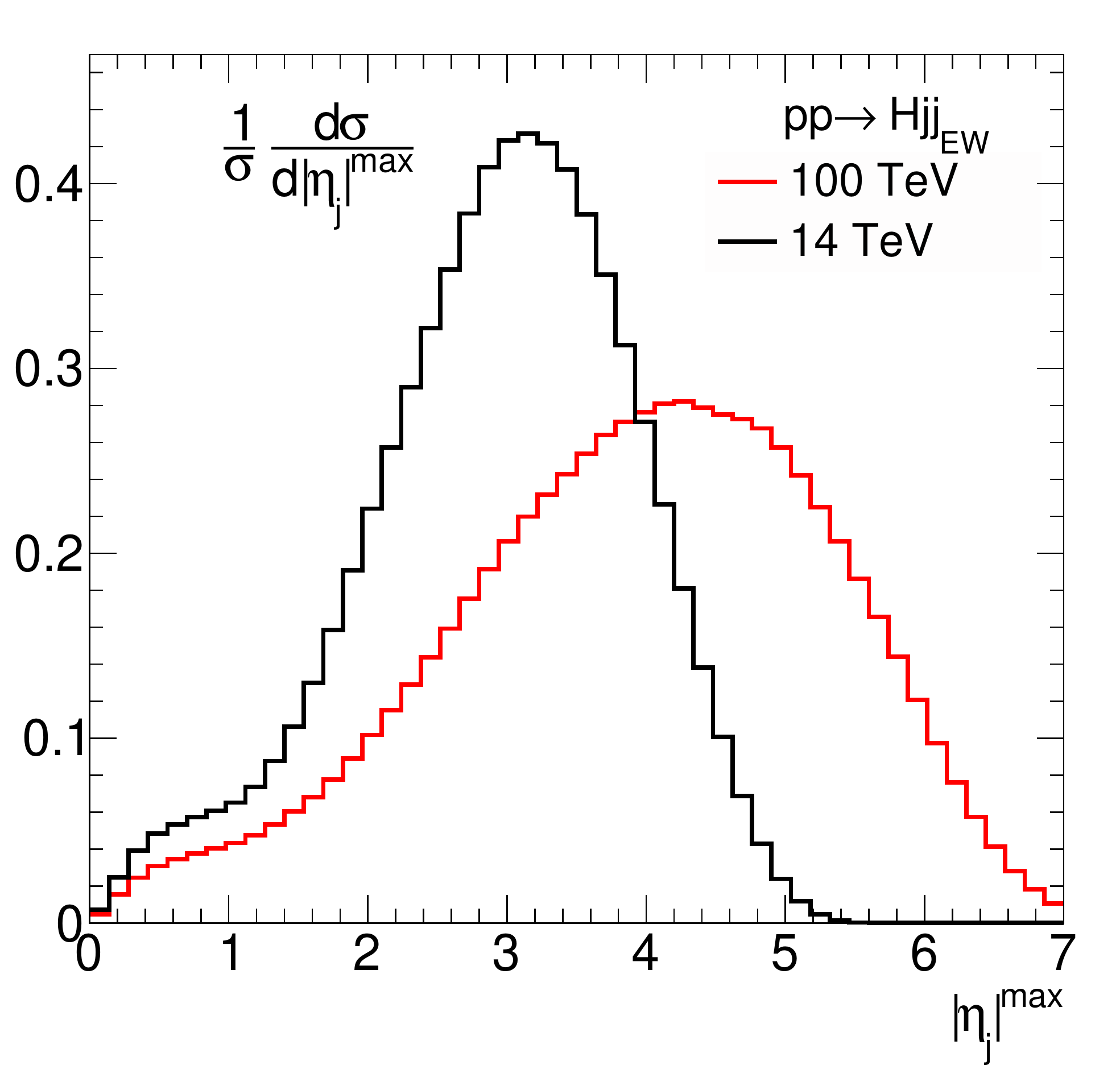}
\hspace{0.1\textwidth}
\includegraphics[width=.35\textwidth]{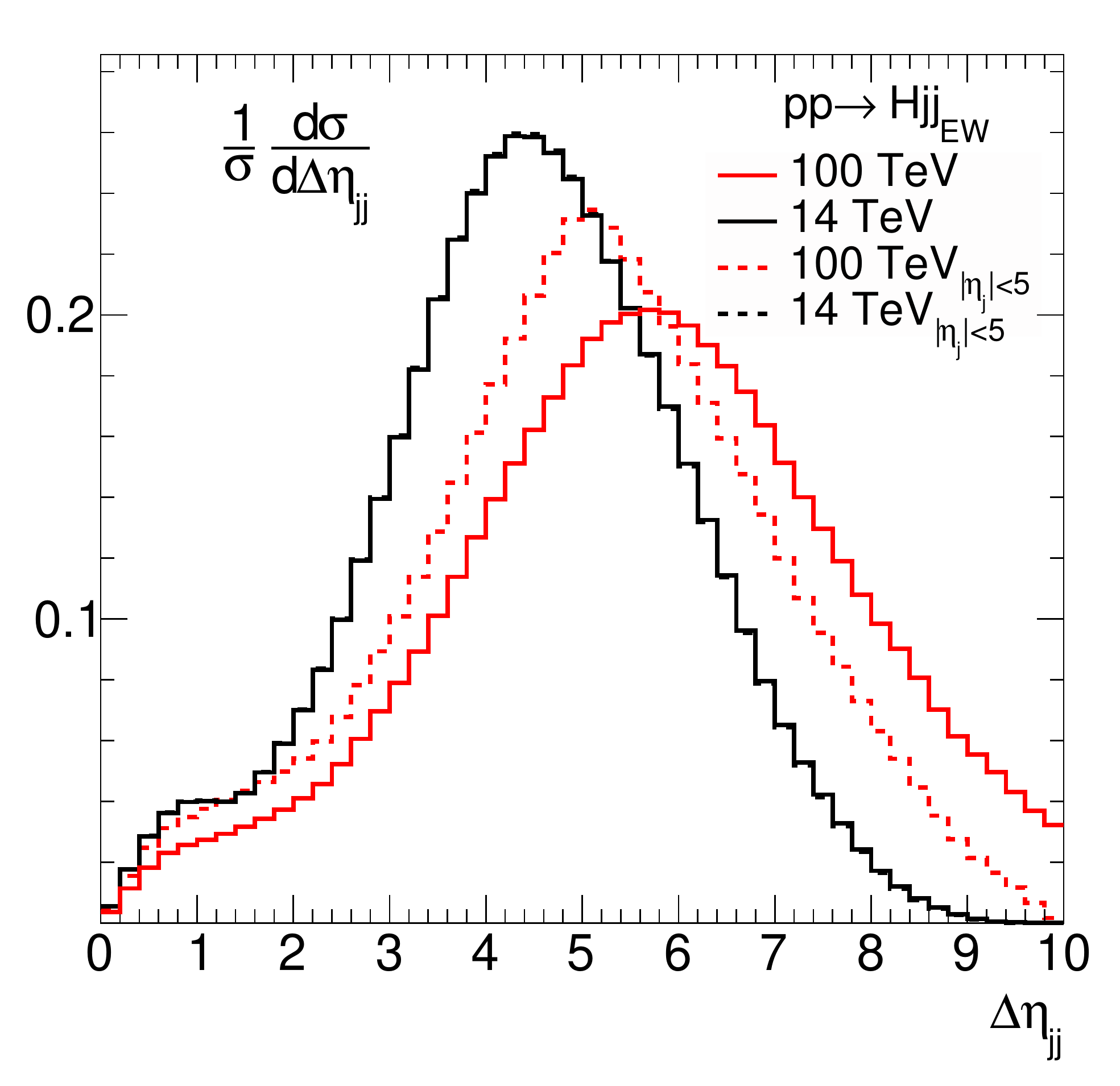}
\caption{Leading tagging jet rapidity $|\eta_j|^\text{max}$ (left) and
  rapidity difference $\Delta\eta_{j_1j_2}$ (right) for the WBF signal
  $Hjj_\text{\text{EW}}$, evaluated at 14~TeV LHC and 100~TeV collider
  energy. We illustrate the effect of the detector cut $|\eta_j|<5$
  on $\Delta \eta_{j_1 j_2}$.}
\label{fig:eta}
\end{figure}

First, we determine what the WBF signal requirements for the two tagging
jets $j_{1,2}$ are~\cite{tagging,scaling}. The signal is
defined by two high-energetic forward jets going into different
hemispheres,
\begin{align}
\eta_{j_1}\cdot\eta_{j_2}<0 \; .
  \label{eq:basic1} 
\end{align}
The effect of increasing the collider energy from 14~TeV LHC to
100~TeV on the more forward jet rapidity and on the rapidity
difference is shown in Fig.~\ref{fig:eta}.  Although it
would be beneficial to extend the detector coverage from $|\eta_j| < 5$
to $|\eta_j| < 6$~\cite{fcc_higgs,hh_wbf}, we only require
\begin{align} 
 |\eta_j|<5
\end{align}
and indicate a possible improvement from extending the detector for
larger rapidities.  In Fig.~\ref{fig:eta}, we observe a shift in the
peak of the rapidity difference from $\Delta \eta_{j_1 j_2} = 4.5$ at
the LHC to $\Delta \eta_{j_1 j_2} = 5$ for a 100~TeV collider with
limited coverage, and to $\Delta \eta_{j_1 j_2} = 5.5$ for all events
at larger energy. We will require
\begin{align}
|\eta_{j_1}-\eta_{j_2}|>5\;,
  \label{eq:basic2} 
\end{align}
instead of the standard choice $\Delta \eta_{j_1 j_2}>4.2$ at the
LHC~\cite{eboli_zeppenfeld}.\medskip

%-------------------------------------------------------
%-------------------------------------------------------

%-------------------------------------------------------
\begin{figure}[b!]
\includegraphics[width=.35\textwidth]{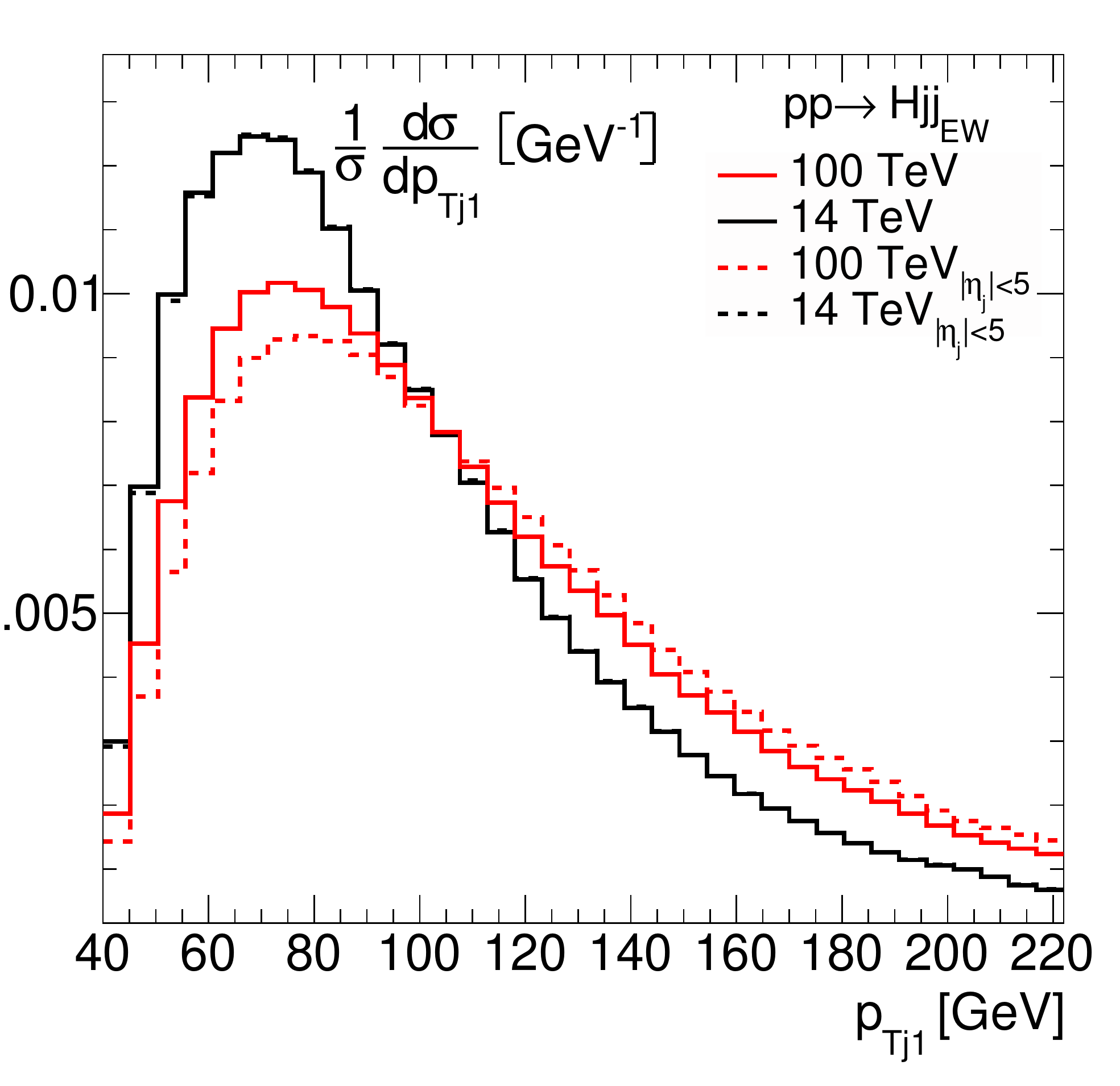}
\hspace{0.1\textwidth}
\includegraphics[width=.35\textwidth]{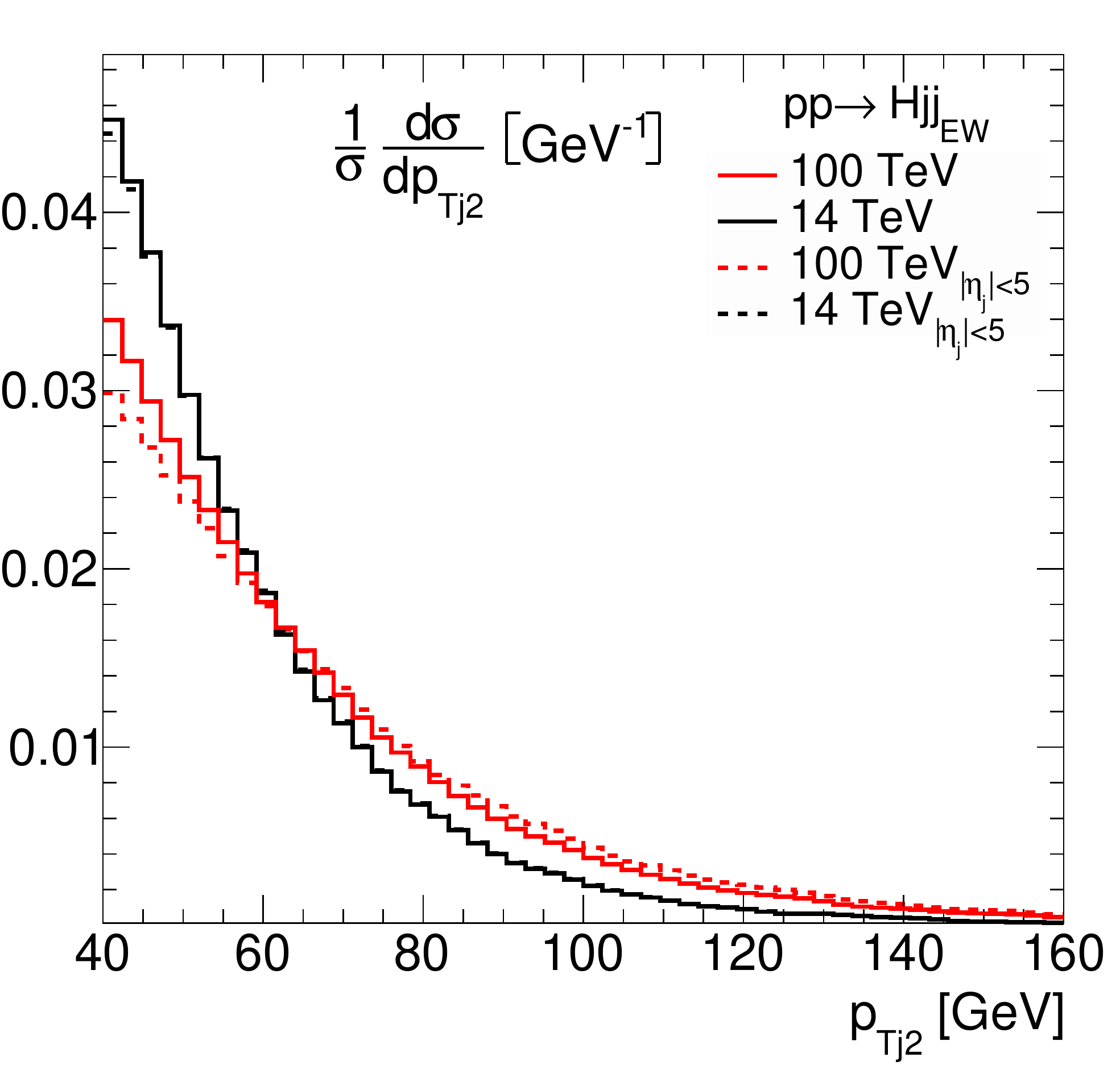}
\caption{Tagging jet transverse momenta $p_{T,j_1}$ (left) and
  $p_{T,j_2}$ (right) for the WBF signal $Hjj_\text{EW}$.}
\label{fig:ptj}
\end{figure}
%-------------------------------------------------------

As the second key observable, we show the transverse momenta of the
tagging jets in Fig.~\ref{fig:ptj}. In contrast to the naive paradigm
that a larger hadronic collider energy leads to more and more
energetic jets from valence quark scattering, we only observe a modest
enhancement on $p_{T,j}$ when we go from LHC energy to 100~TeV.  The
reason for this is that the typical tagging jet transverse momentum is set by
the $W$ and $Z$-masses and the massive gauge boson splitting kernel.
Given a quark with energy $E$, the probability of finding a collinear
jet-boson pair with a boson energy $xE$ and a transverse momentum
$p_T$ is given by~\cite{splitting1,splitting2}
\begin{align}
P_T(x,p_T) &\propto 
                 \frac{1+(1-x)^2}{x} \;
                 \frac{p_T^3}{(p_T^2 + (1-x) \, m_V^2)^2}  \; ,\notag \\
P_L(x,p_T) &\propto 
                 \frac{(1-x)^2}{x} \;
                 \frac{m_V^2 p_T}{(p_T^2 + (1-x) \, m_V^2)^2} \; ,
\label{eq:splitting}
\end{align}
where $T~(L)$ stands for the transverse (longitudinally) polarized
gauge boson $V$. While for $p_T\ll m_V$ the transverse splitting
probability, $P_T(x)$, is suppressed, for $p_T\gg m_V$ the longitudinal
splitting probability, $P_L(x)$, decreases faster with increasing $p_T$.  
Altogether, this means that
kinematic differences between the LHC and a 100~TeV collider are 
largely limited to the tagging jet rapidities.\medskip

%-------------------------------------------------------
\begin{figure}[t]
\includegraphics[width=.32\textwidth]{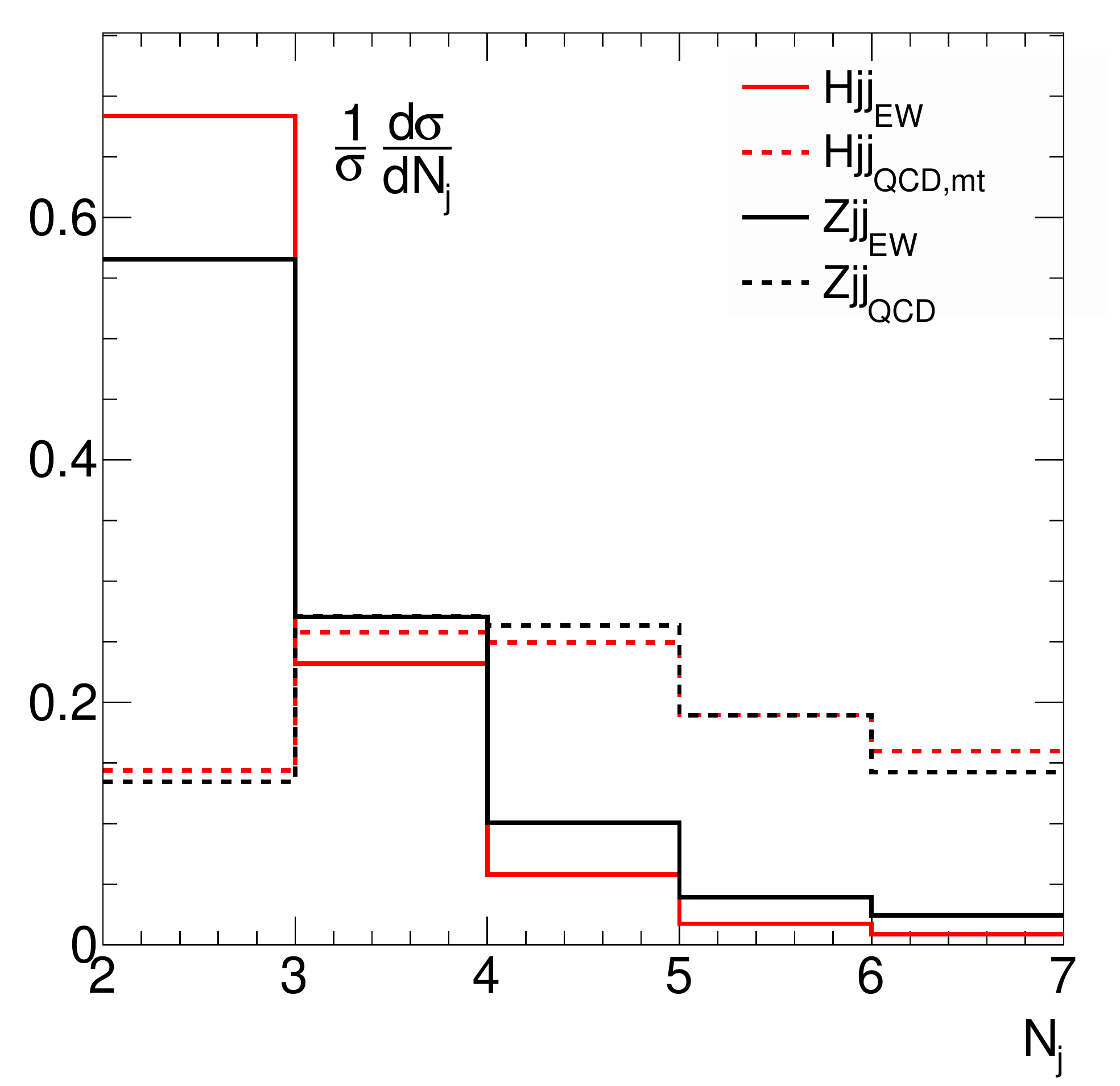}
\includegraphics[width=.32\textwidth]{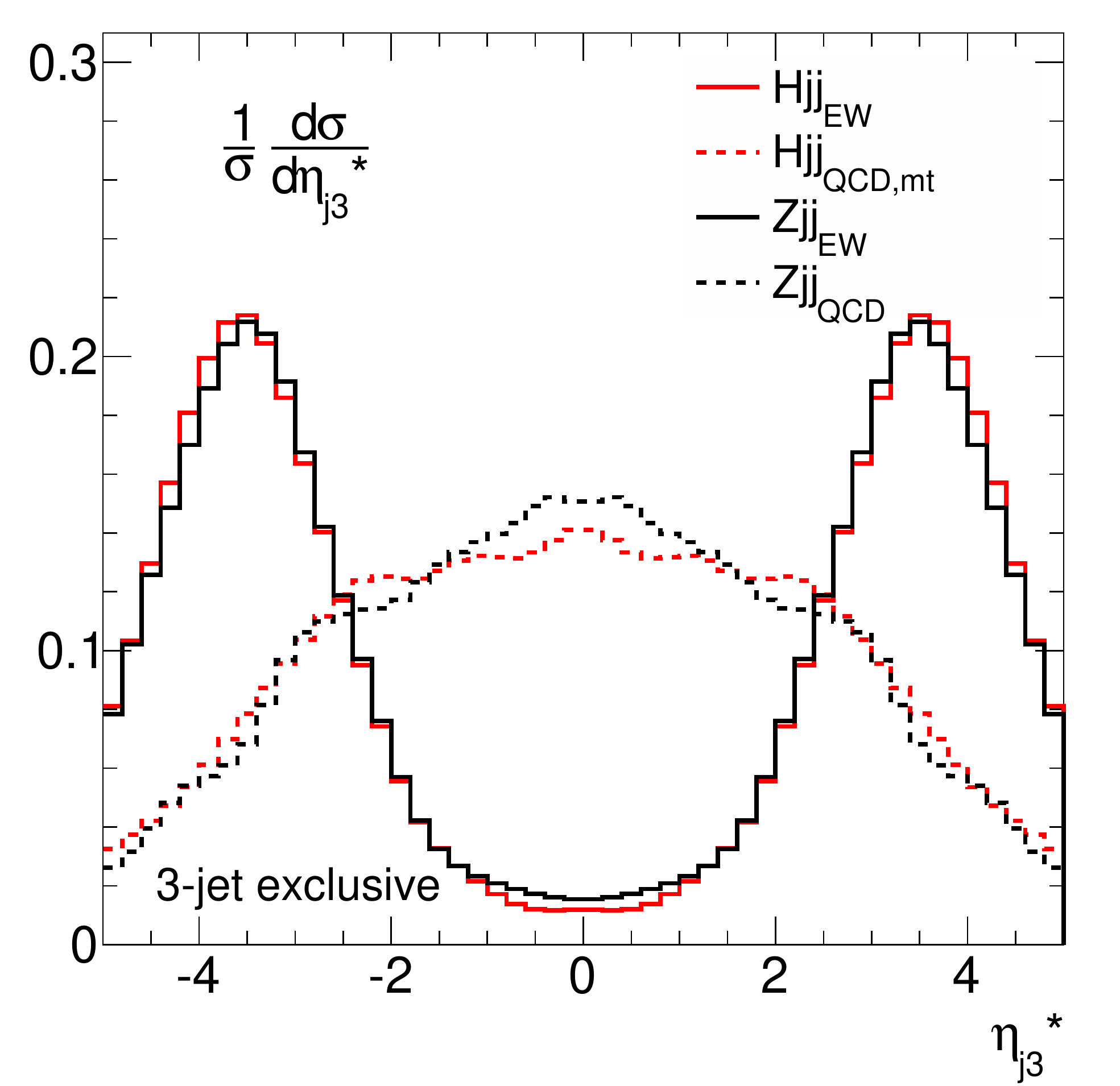}
\includegraphics[width=.32\textwidth]{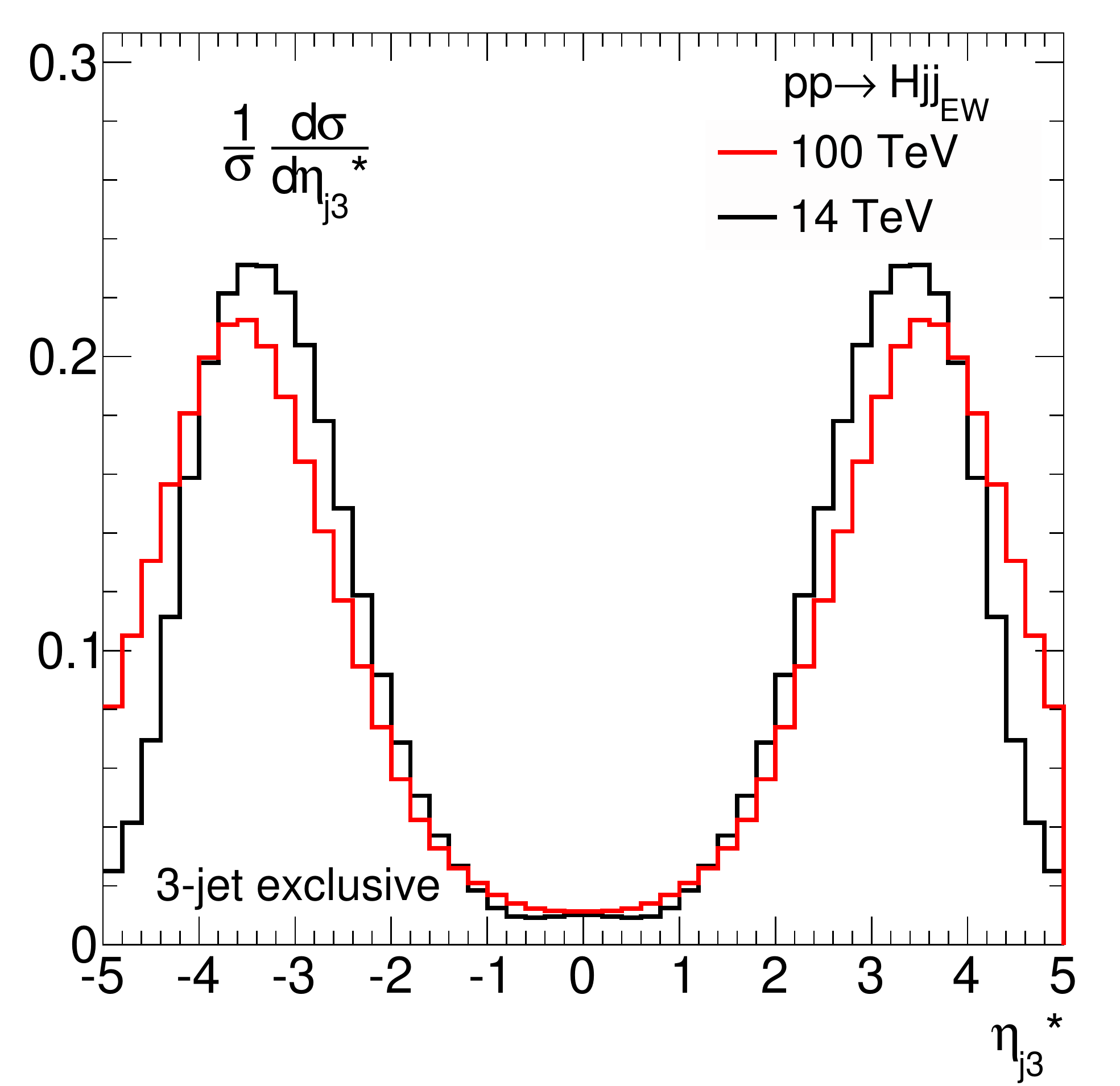}
\caption{Exclusive number of jets (left) and $\eta_{j_3}^*$ (central
  and right) distributions for the WBF and the two $Zjj$
  backgrounds. In the right panel we do not apply any cut on
  $|\eta_{j_{1,2}}|$.}
\label{fig:njet}
\end{figure}
%------------------------------------------------------

The defining feature of any WBF signal at hadron colliders is the
suppressed central jet radiation, as compared to the QCD processes
illustrated in Fig.~\ref{fig:feyn}~\cite{cjv}. Beyond the usual
perturbative QCD arguments, the fundamental reason is the different
Poisson vs staircase pattern in the number of radiated
jets~\cite{staircase}. In the left panel of Fig.~\ref{fig:njet}, we see
how this leads to a reduced jet activity in the signal and can be
exploited by a simple jet veto to enhance the signal-to-background
ratio to the level of all-electroweak signal and background processes.
The samples assume stable Higgs and $Z$-bosons, and all jets are defined
as anti-$k_T$ jets~\cite{fastjet} with $R=0.4$, $p_{T,j}>20$~GeV,
and $|\eta_j|<5$.  The WBF event selection includes Eqs.\eqref{eq:basic1}
and~\eqref{eq:basic2}, as well as 
\begin{align} 
p_{T,j_{1,2}}>40~\gev \qqquad \text{and} \qquad 
m_{j_1 j_2}>1200~\gev \; ,
  \label{eq:basic3} 
\end{align} 
for the two tagging jets.  An obvious question is how much a dedicated
analysis of the third jet kinematics can improve over the simple
veto~\cite{cathy}. We require this third jet to have a minimum
transverse momentum with a default value of 
\begin{align}
p_{T,j_3} > p_{T,\text{veto}} = 20~\gev \; .
\label{eq:pveto1}
\end{align} 
In Fig.~\ref{fig:njet} we also show the relevant kinematic variable,
$\eta_{j_3}^*$, for which we require
\begin{align}
\eta_{j_3}^*=\eta_{j_3}-\frac{\eta_{j_1}+\eta_{j_2}}{2} > 3 \; .
  \label{eq:eta3} 
\end{align}
While the electroweak signal and backgrounds show a suppression for
$\eta_{j_3}^*=0$, the QCD background and QCD Higgs production are
centered there. We explore these features by defining 2-jet and 3-jet
samples by
\begin{enumerate}
\item vetoing a third jet for $p_{T,j_3} > p_{T,\text{veto}}$;
\item requiring a third jet with $p_{T,j_3} > p_{T,\text{veto}}$ and
  $|\eta_{j_3}^*|>3$, vetoing a fourth jet for $p_{T,j_4} >
  p_{T,\text{veto}}$.
\end{enumerate}
For this two-step strategy it is crucial that we order the jets
according to their transverse momenta, \ie the two hardest jet
fulfilling our tagging jet criteria are marked as tagging jets. A
third, softer jet can then, in principle, be more forward than either of
the the tagging jets.  Interestingly, the suppression of the central jet activity is
even more pronounced at 100~TeV than at the LHC, as shown in the right
panel of Fig.~\ref{fig:njet}.  Following Fig.~\ref{fig:eta} the
tagging jets have larger rapidities at larger collider energies, so
the third jet will follow the tagging jets towards larger
rapidities.\medskip

\begin{figure}[t]
 \includegraphics[width=0.34\textwidth]{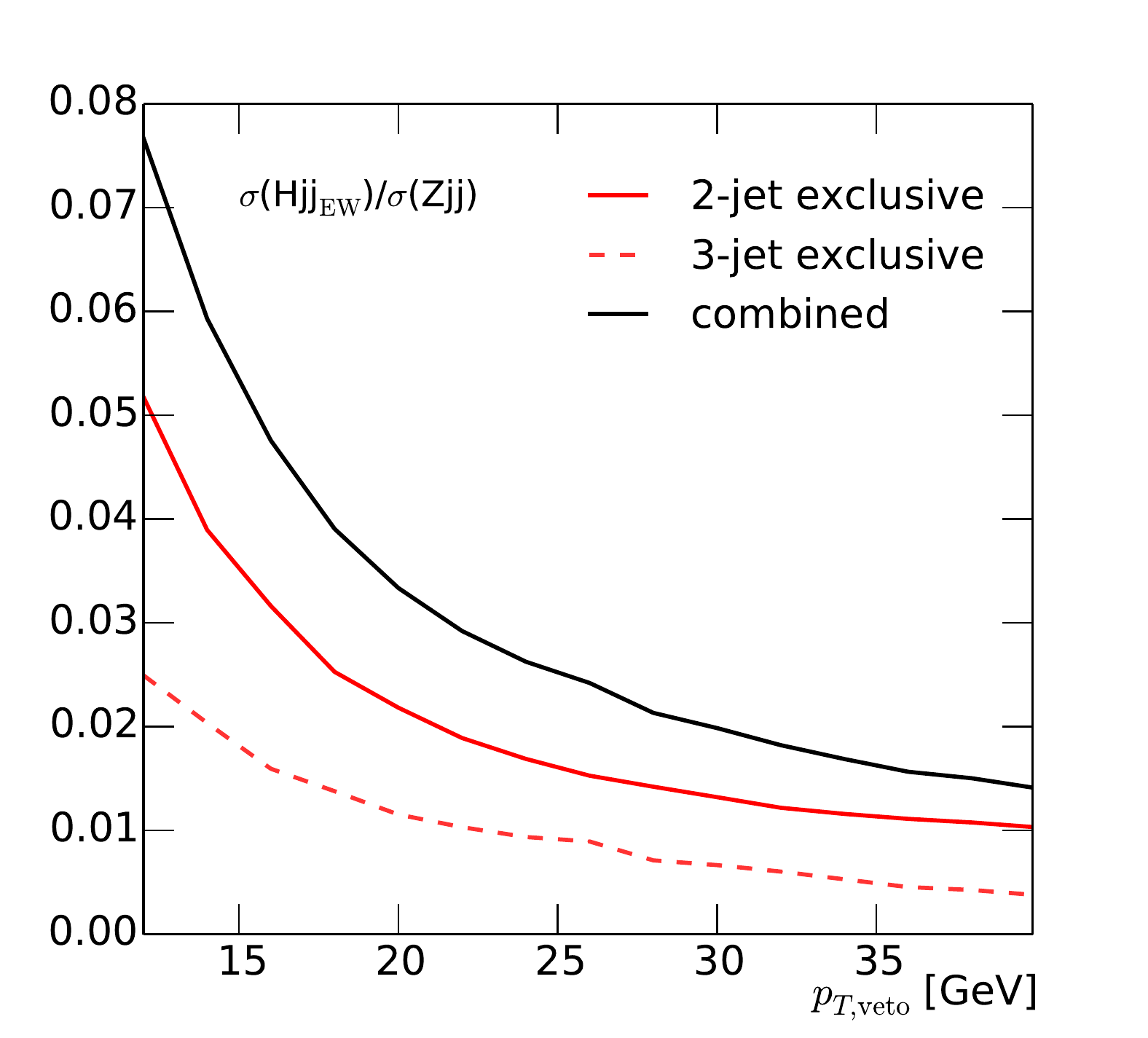}
 \hspace{-0.025\textwidth}
 \includegraphics[width=0.34\textwidth]{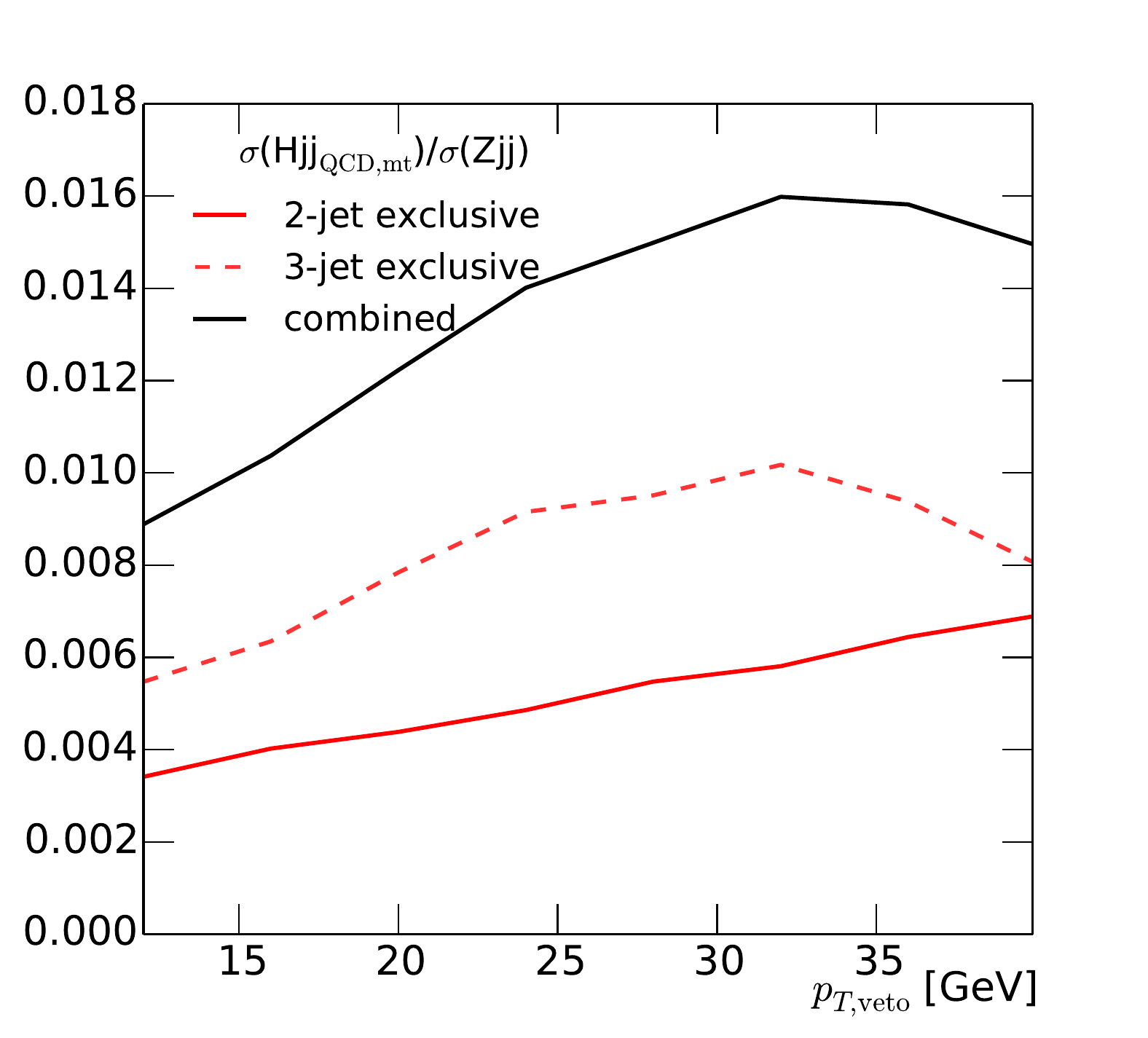}
 \hspace{-0.025\textwidth}
 \includegraphics[width=0.34\textwidth]{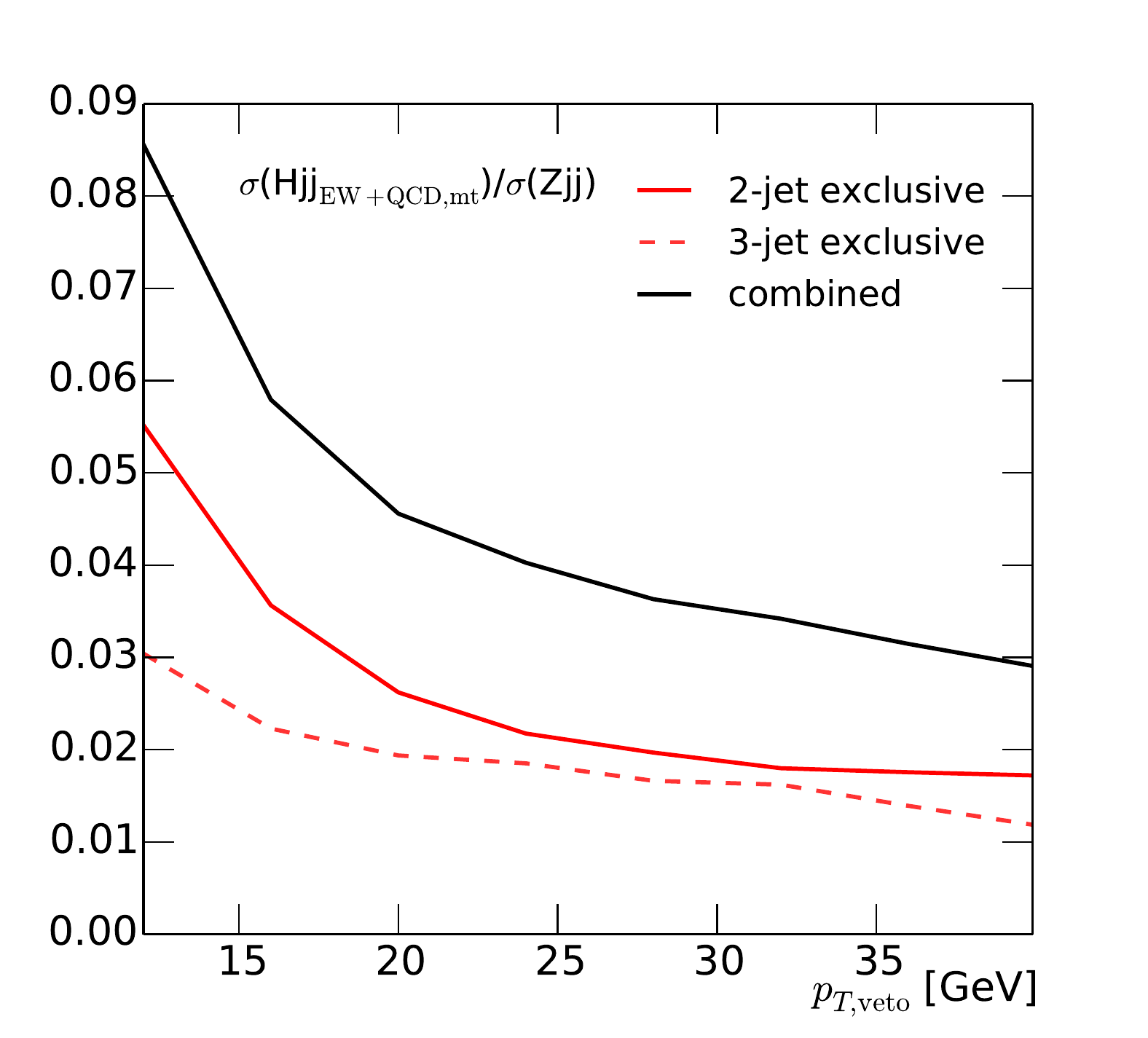}
  \caption{Signal-to-background ratio for WBF production (left), gluon-fusion
    production (center), and the combined Higgs signal (right) vs
    dominant QCD and EW $Zjj$ backgrounds. The jet multiplicities are
    exclusive.}
\label{fig:significance}
\end{figure}

The scale above which a fourth jet is vetoed clearly impacts the
signal-to-background ratio and on the significance of any Higgs
signal. The expectation is for $S/B$ and $S/\sqrt{B}$ to decrease with
increasing veto scale, because the QCD-dominated background and signal
benefits largely from the increased phase space for additional
radiation, whereas the WBF signal does not.
Figure~\ref{fig:significance} shows the signal-to-background ratio for
both, the gluon fusion and WBF channels, against the combined QCD and
EW $Zjj$ backgrounds. For the WBF signal the ratio decreases as the
veto scale is increased~\cite{scaling}. This behavior can be seen in
both the 2-jet and the 3-jet channels. The gluon fusion channel
behaves very differently. Because it is a QCD process with Poisson
scaling~\cite{scaling}, increasing the veto scale increases the
sensitivity of the channel. This is a linear increase with the veto
scale for the 2-jet channel, while the contribution from the 3-jet
channel begins to decrease as it approaches high veto scales.  This is
because the two leading jets are required to have $p_T>40$~GeV, so as
the third jet $p_T$ is forced to approach this limit there is a
reduced phase space for its emission.  At this point, with a veto on
the fourth jet above $p_{T,j} > 40$~GeV, the signal sample consists of
1/3 gluon fusion events, even after WBF cuts. 

%------------------------------------------------------
%------------------------------------------------------

%------------------------------------------------------
\begin{figure}[b!]
 \begin{center}
 \includegraphics[width=0.45\textwidth]{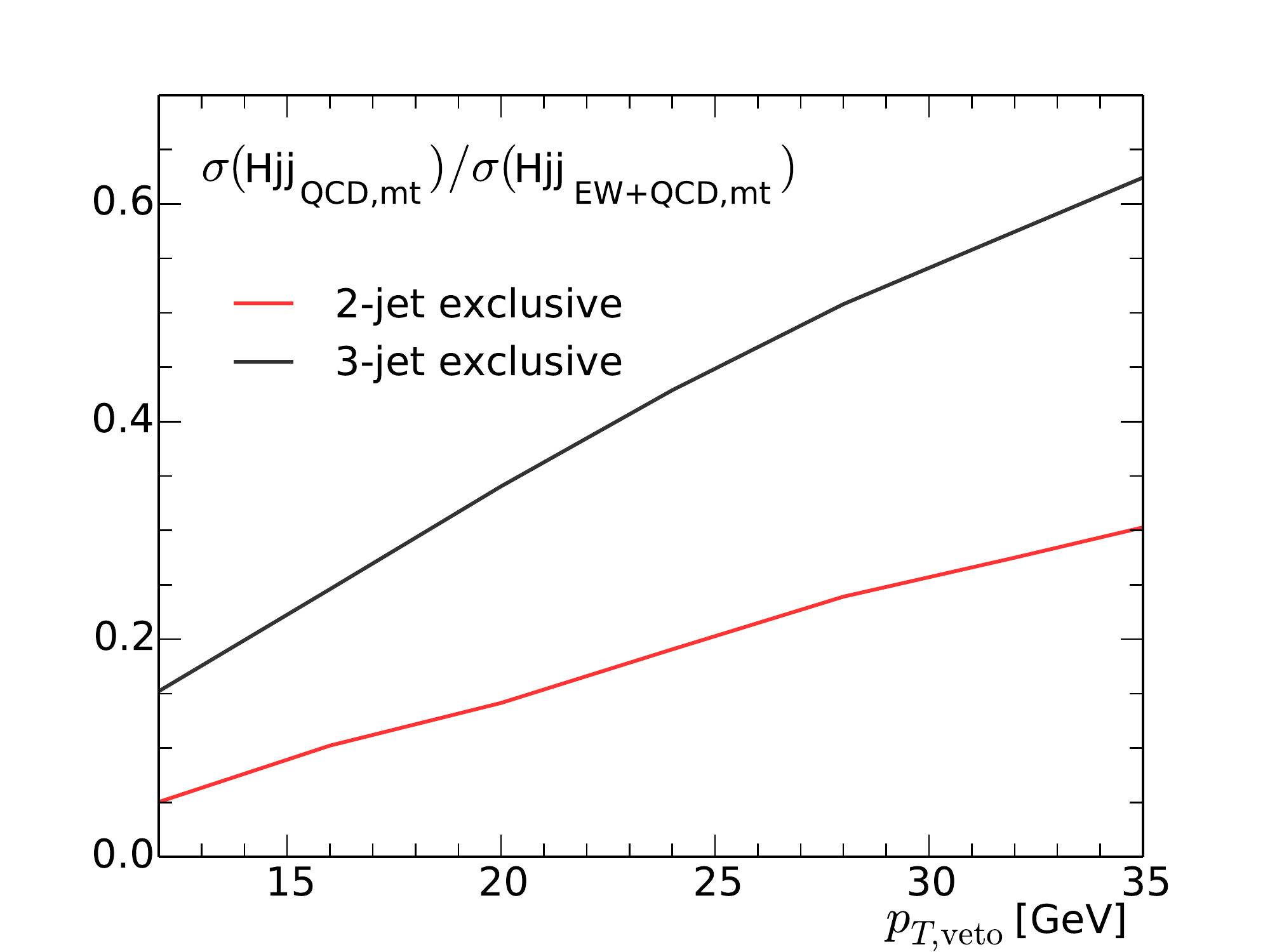}
 \end{center}
  \caption{Fractions of gluon fusion events in the Higgs signal after
    WBF cuts, as a function of the veto scale for the fourth jet.}
\label{fig:fractions}
\end{figure}
%------------------------------------------------------

In the right panel of Fig.~\ref{fig:significance} we show the
signal-to-background ratio for the combined Higgs signal. The 2-jet
channel shows a flattening behavior at veto scales above 30~GeV.  The
3-jet channel behaves more in line with expectations, and the
signal-to-background ratio slowly decreases with veto scale. Combining
the two channels leads to a decrease in the ratio in the range we
study.  Throughout this paper we use a default veto scale of
$p_{T,\text{veto}} = 20~\gev$.  This choice of veto scale is identical
to the scale which separates the 2-jet and 3-jets samples, as defined
in Eq.\eqref{eq:pveto1}.  If the $p_T$ requirement for the tagging
jets were higher and the 3-jet significance for the gluon fusion
channel did not fall off so quickly, the sensitivity in the high veto
scale region could even increase. This effect is also present, and to
a slightly larger extent, in a signal over square-root background 
analysis.

In Fig.~\ref{fig:fractions} we show how the fraction of gluon fusion
events in the Higgs signal changes with the jet veto scale, after
applying the usual WBF cuts. The range of veto scales is limited by
the tagging jet requirement $p_{T,j_{1,2}} > 40$~GeV.  The exclusive
3-jet rate with our two-step veto strategy dominates the analysis
power and for realistic veto scales of $p_{T,\text{veto}} =
20~...~35$~GeV, the gluon fusion contamination varies between 15\%
and 30\% for the 2-jet channel and between 35\% and 60\% for the 3-jet
channel. This kind of analysis should eventually allow us to reduce
our dependence on Monte Carlo predictions and separate the two Higgs
production processes based on data.

%%%%%%%%%%%%%%%%%%%%%%%%%%%%%%%%%%%%%%%%%%%%%%%%%%
\section{Higgs to invisibles}
\label{sec:inv}

%-------------------------------------------------------
\begin{figure}[b!]
\includegraphics[width=.42\textwidth]{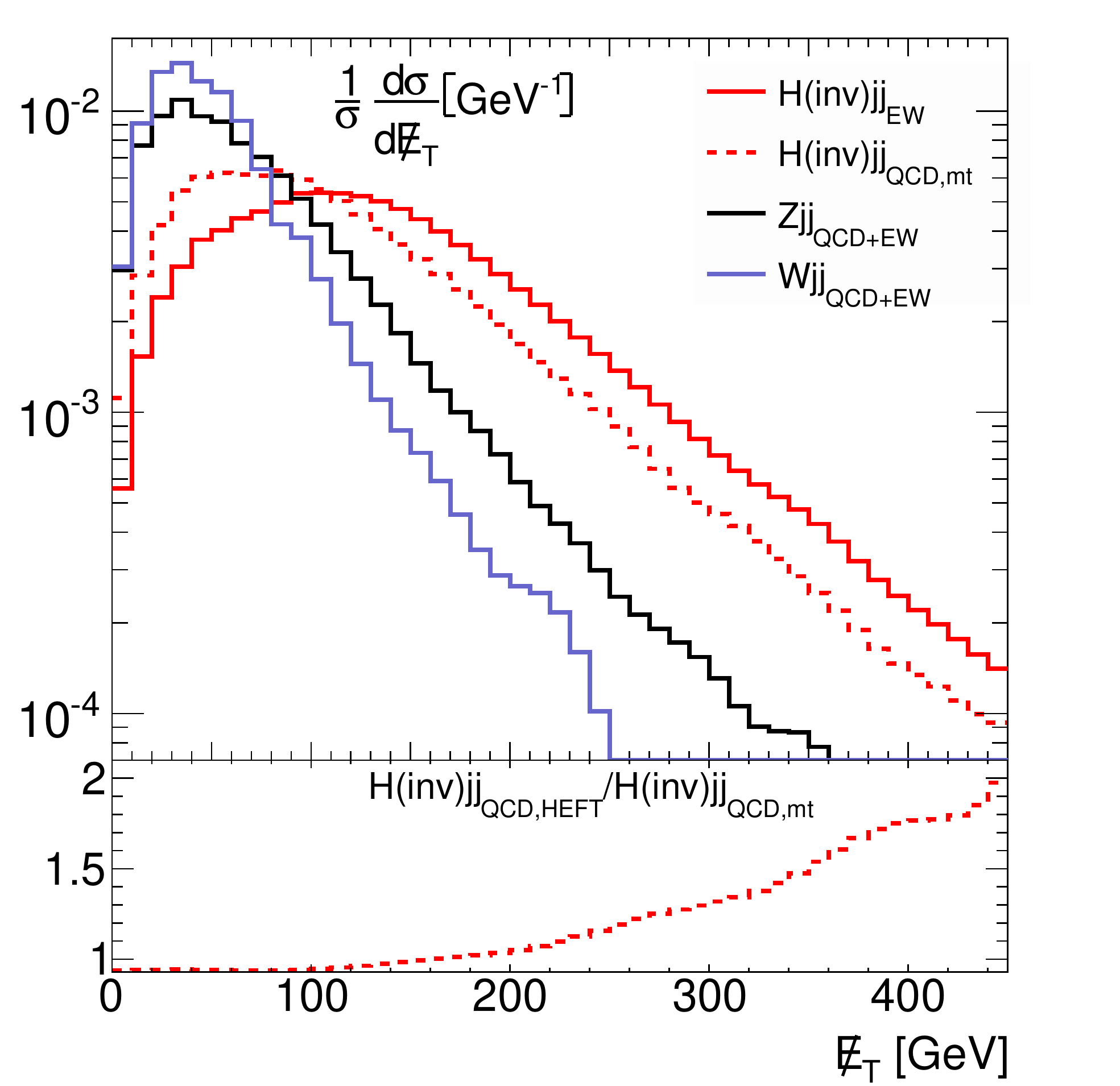}
\caption{Missing transverse energy distribution $\met$ for
  signal and backgrounds.  The bottom panel displays the ratio between
  the heavy-top approximation and the correct result for the QCD Higgs
  production process.}
\label{fig:met}
\end{figure}
%-------------------------------------------------------

For a first analysis making use of tagging jets at a 100~TeV collider,
we turn to Higgs decays to invisible
particles~\cite{eboli_zeppenfeld,cathy}, where the corresponding
branching ratio $\text{BR}(H\to \text{inv})$ is part of modern global
Higgs coupling analyses~\cite{legacy}.  The main backgrounds are $Zjj$
and $Wjj$ production, as illustrated in Fig.~\ref{fig:feyn}.  At the
level of our analysis detector effects, aside from acceptance cuts,
should not play any noticeable role. The only exception is the missing
transverse momentum measurement, for which we include a gaussian
smearing of $\Delta\met=20~\gev$.\medskip

We start our analysis requiring two tagging jets with
\begin{align}
\eta_{j_1}\cdot\eta_{j_2}<0 
\qqquad 
p_{T j_{1,2}}>40~\gev 
\qqquad 
|\eta_{j_{1,2}}|<5
\qqquad 
\Delta \eta_{j_1 j_2} > 5 \; .
\label{eq:ptj} 
\end{align}
Following the original analysis~\cite{eboli_zeppenfeld}
we apply an additional cut on the azimuthal angle between the tagging jets
\begin{align}
\Delta\phi_{j_1 j_2}<1\;,
  \label{eq:dphijj} 
\end{align}
which is sensitive to the Lorentz structure of the hard
interaction~\cite{phi_jj}.

%-------------------------------------------------------
\begin{figure}[t]
\includegraphics[width=0.35\textwidth]{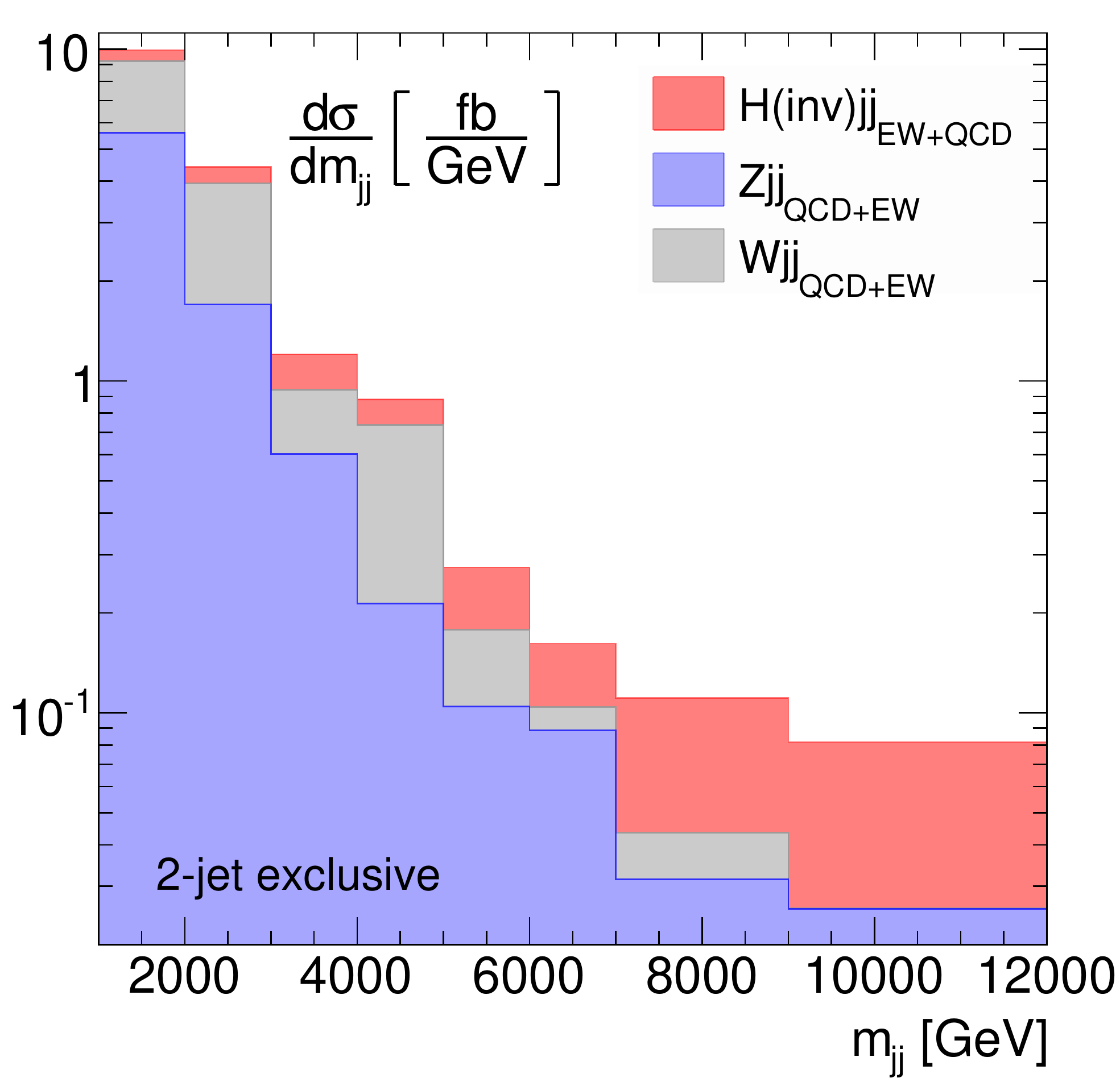}
\hspace{0.1\textwidth}
\includegraphics[width=0.35\textwidth]{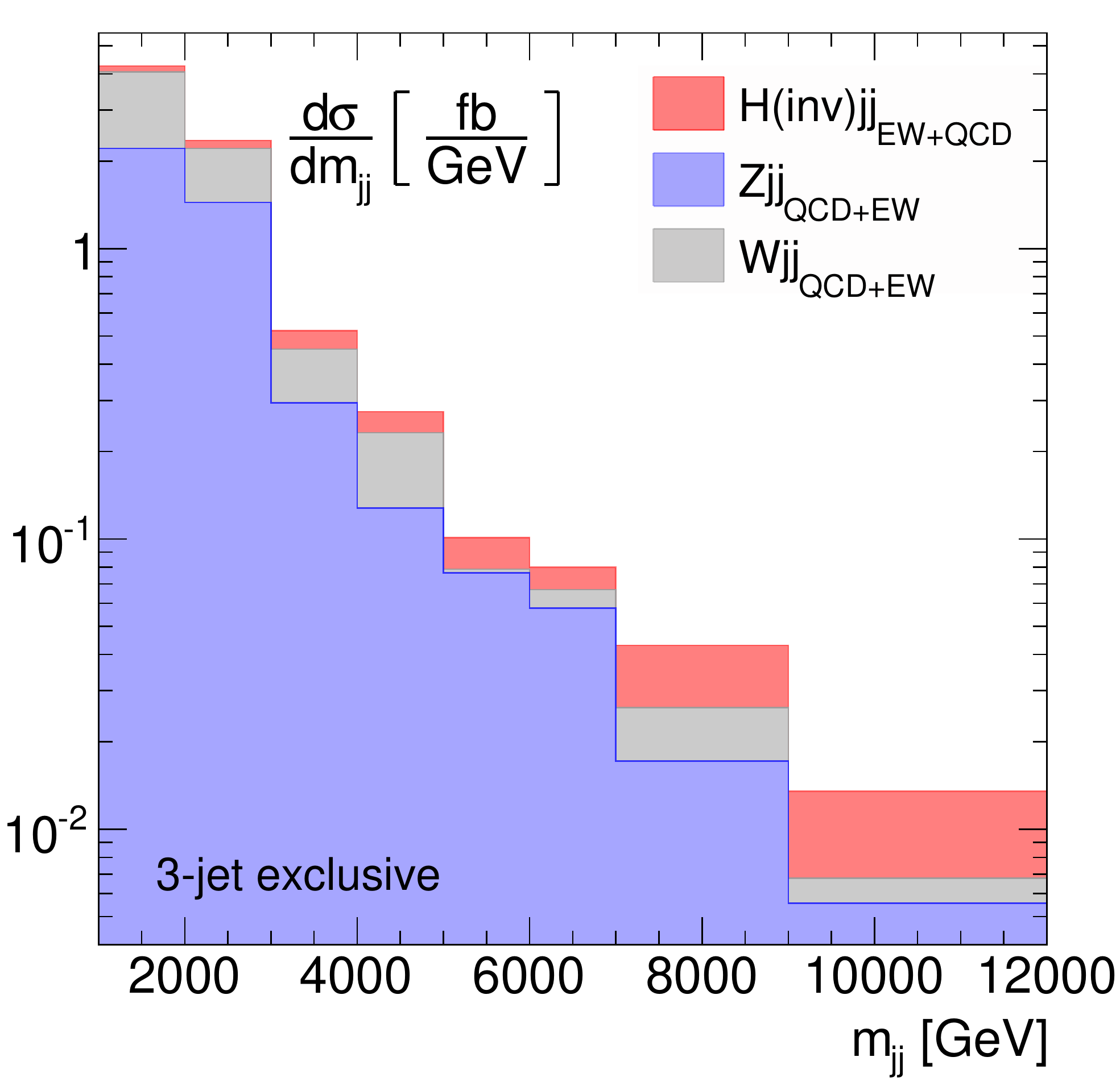}
\caption{Stacked invariant mass distribution $m_{j_1 j_2}$ for the two
  tagging jets, separated for the exclusive 2-jet (left) and
  3-jet (right) samples.}
\label{fig:mjj}
\end{figure}
%-------------------------------------------------------

Our two-step central jet veto is based on Eqs.\eqref{eq:pveto1}
and~\eqref{eq:eta3}.  In addition, we veto any isolated lepton where
the isolation criterion requires less than 20\% of hadronic activity
in a radius $R=0.2$ around the lepton. This requirement is shown to be
very efficient in suppressing the $W$+jets background.\medskip

In Fig.~\ref{fig:met}, we display the normalized $\met$ distributions
for the two Higgs channels and the backgrounds.  In the bottom panel
we show how the heavy-top approximation for this channel fails towards
large missing energies $\met \gtrsim
m_t$~\cite{hjets1,hjets2}. Clearly, any measurement including a
missing transverse energy cut needs to account for the full top mass
dependence.  Because both backgrounds peak around $\met = 40$~GeV and the gluon
fusion Higgs channel peaks around $\met = 60$~GeV, for the
extraction of the WBF signal we require
\begin{align}
\met>100~\gev \; .
  \label{eq:met} 
\end{align}
\medskip

%-------------------------------------------------------
\begin{figure}[t]
  \includegraphics[width=0.40\textwidth]{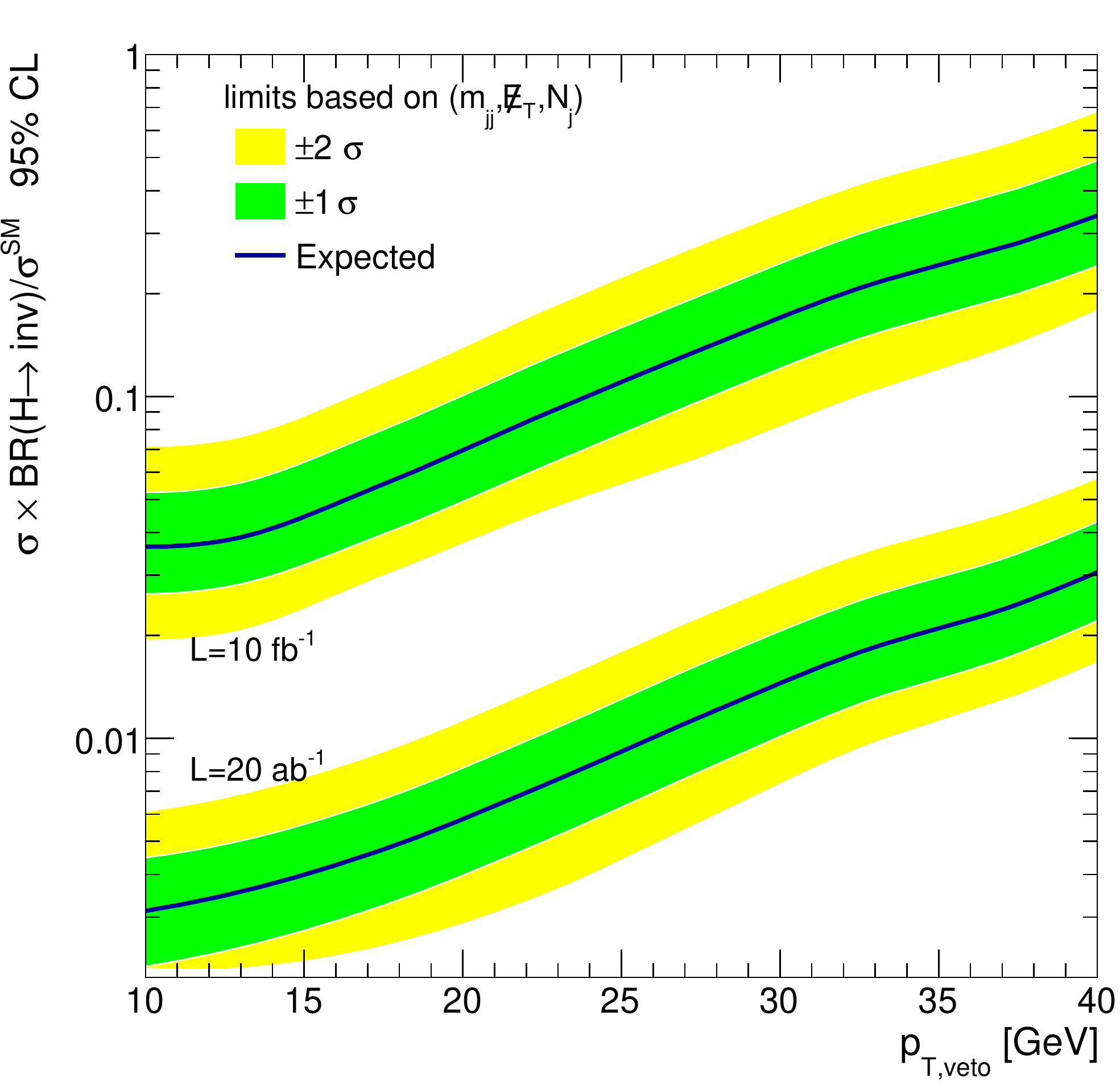}
  \hspace{0.1\textwidth}
  \includegraphics[width=0.40\textwidth]{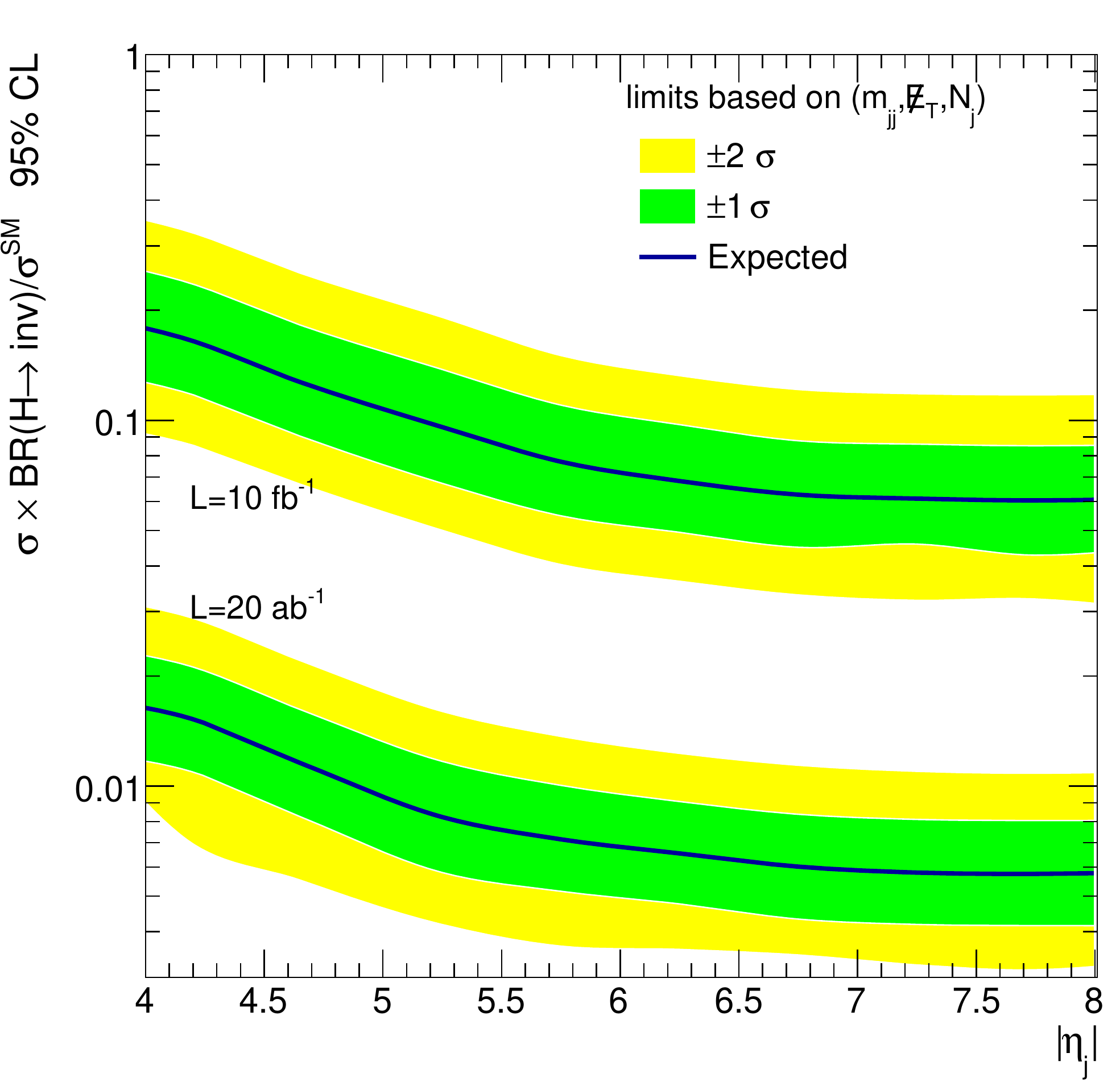}
\caption{Expected 95\% CL bound on the invisible Higgs branching
  ratio, based on a log-likelihood analysis of the three-dimensional
  distribution $\{ \met,m_{j_1 j_2},N_j \}$.  We vary the  $p_{T,\text{veto}}$
  with ${p_{T,j_3}>p_{T,\text{veto}}}$  (left) and the maximum rapidity
  $|\eta_j|$ (right).}
\label{fig:cls_ptj}
\end{figure}
%-------------------------------------------------------

After applying all cuts we arrive at the invariant mass distributions
of the tagging jets shown in Fig.~\ref{fig:mjj}. While we have
confirmed that the transverse momentum spectrum does not significantly
change when we go from the LHC to 100~TeV, this is obviously not true for
the longitudinal momenta or the invariant mass of the tagging
jets. The regime sensitive to the WBF Higgs signal at 100~TeV starts
only around $m_{j_1 j_2} \gtrsim 7$~TeV, indicated by a
signal-to-background ratio around one. For our estimate of the
collider reach we rely on the three kinematic observables
\begin{align}
\{ \, \met,m_{j_1 j_2},N_j \, \} \; ,
\label{eq:loglikely_inv}
\end{align}
within their allowed range $\met > 100$~GeV and $N_j = 2,3$. They are
chosen to include information on the tagging jets ($m_{j_1 j_2},N_j$),
as well as everything we know about the Higgs momentum ($\met$).

To estimate the constraining power on $\text{BR}(H\to \text{inv})$, we
perform a three-dimensional binned log-likelihood analysis for
$\textsc{CL}_s$ based on the vector of kinematic distributions shown
in Eq.\eqref{eq:loglikely_inv}. It exploits the rate and the shape
information in the two panels of Fig.~\ref{fig:mjj}, combined with the
$\met$ dependence.  For an early running with an integrated luminosity
of $10~\ifb$ we use the systematic uncertainties from the CMS mono-jet
search~\cite{Khachatryan:2014rra}, where the systematics for
$Z\rightarrow\nu\nu$ and ${W\rightarrow \ell \nu}$ backgrounds range
around 5\% on the background rate. These uncertainties are modeled as
nuisance parameters.  For the target luminosity of $20~\iab$ we
assume these uncertainties to reach 0.5\%, hoping for a better
understanding of the systematic uncertainties with more data. This is
significantly worse than the scaled luminosity would suggest, so all of
our results will be systematics limited.\medskip

In the left panel of Fig.~\ref{fig:cls_ptj} we show the expected
95\%~CL bound on $\text{BR}(H\to \text{inv})$ as a function of the
minimum transverse energy of the third jet, keeping the two tagging
jets at $p_{T,j_{1,2}}>40~\gev$ and the detector coverage at
$|\eta_j|<5$. A reduced third jet threshold will enhance the effect of
our analysis of the third jet kinematics~\cite{cathy}.  Even using
track jets or even objects without a jet reconstruction, we do not
expect to be able to go below 10~GeV, because of underlying event and
pile-up. The experimental challenge in searching for invisible Higgs
decays turns out to be the same as for dark matter
searches~\cite{DM100TeV}; we need to increase the collider energy
while at the same time keeping the detector thresholds as low as
possible.

In the right panel of Fig.~\ref{fig:cls_ptj}, we estimate the impact
of the rapidity coverage for the two tagging jets. The threshold for
the third jet is kept at $p_{T,j_3} > 30$~GeV.  As expected from
Fig.~\ref{fig:eta}, there are only minor gains if we extend the
detector range past $|\eta_j|\sim 6$. Altogether, we find that with an
excellent detector performance and similarly good control of the
systematics a reach of
\begin{align}
\text{BR}(H \to \text{inv}) \lesssim 0.5\%
\label{eq:bottom_inv}
\end{align}
for a Standard Model production rate appears to be realistic at a
100~TeV machine. However, as mentioned before, any number below one
per-cent strongly relies on our assumptions on background systematics.

%%%%%%%%%%%%%%%%%%%%%%%%%%%%%%%%%%%%%%%%%%%%%%%%%%
\section{Higgs to muons}
\label{sec:muons}

%-------------------------------------------------------
\begin{figure}[t]
\includegraphics[width=0.35\textwidth]{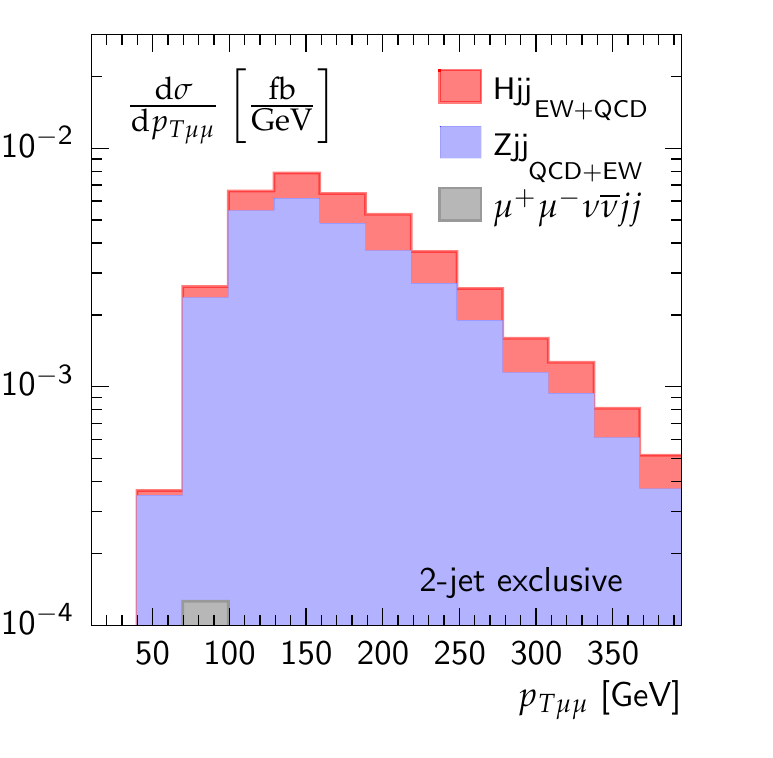}
\hspace{0.1\textwidth}
\includegraphics[width=0.35\textwidth]{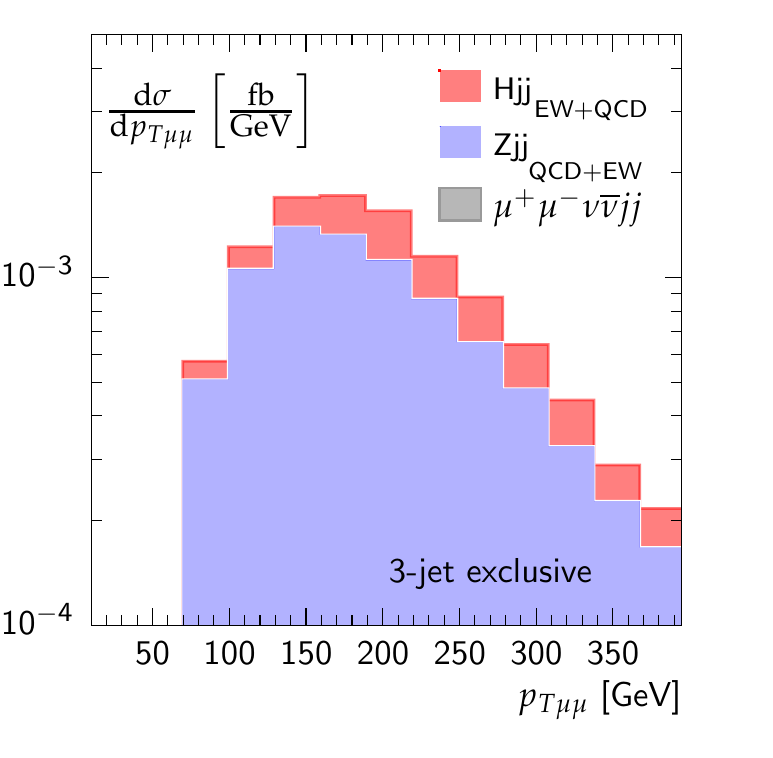}
\caption{Stacked transverse momentum distribution for the $\mu \mu$
  system for the exclusive 2-jet (left) and 3-jet (right)
  samples. We require an invariant mass window
  $|m_{\mu\mu}-m_H|<5$~GeV.}
\label{fig:mjj_mu}
\end{figure}
%-------------------------------------------------------

As a second benchmark process to illustrate WBF Higgs production at
100~TeV, we consider the decay to second-generation leptons
$H\rightarrow\mu^+\mu^-$. This decay channel will barely be observable
at the LHC, and will hardly lead to a precise measurement of the muon
Yukawa coupling in the Standard Model~\cite{lhc_muons}.  The dominant
$Zjj$ backgrounds are illustrated in Fig.~\ref{fig:feyn}.  In
addition, we consider off-shell di-boson and $t\bar{t}$ backgrounds,
denoted as $\mu^+\mu^-\nu\overline{\nu}jj$. Their effect is very
small.\medskip

%-------------------------------------------------------
\begin{figure}[b!]
\includegraphics[width=0.495\textwidth]{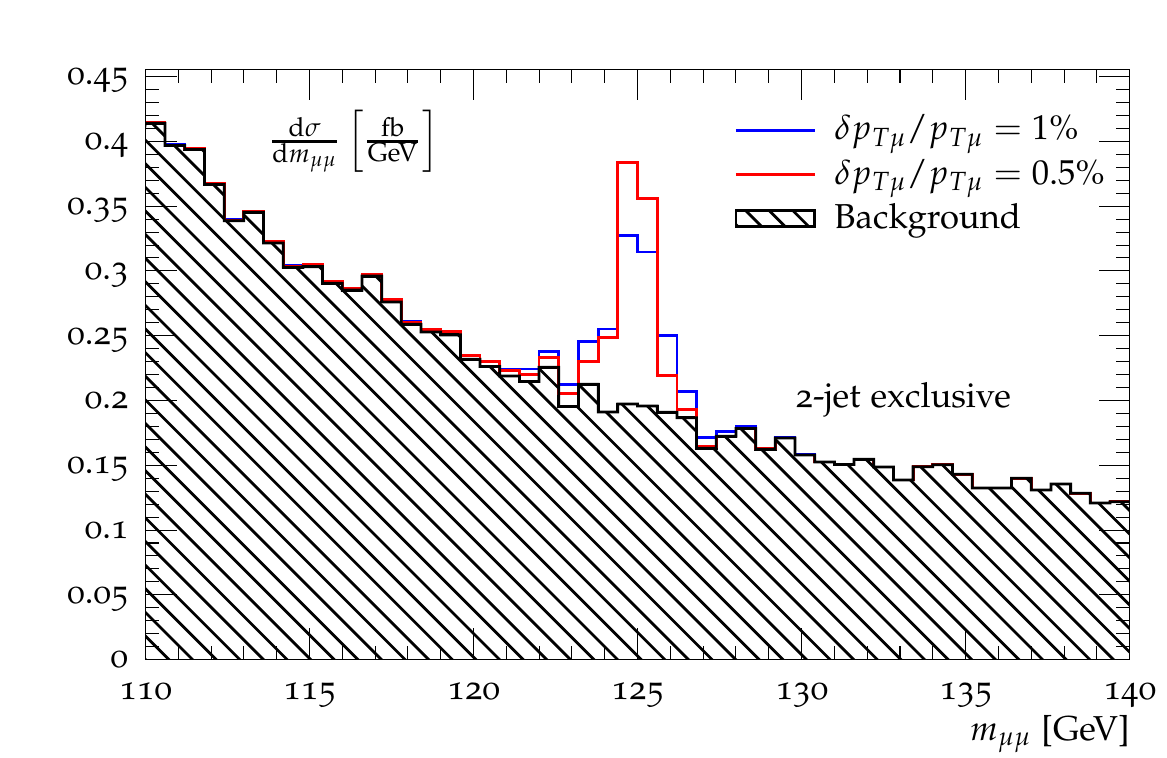}
\includegraphics[width=0.495\textwidth]{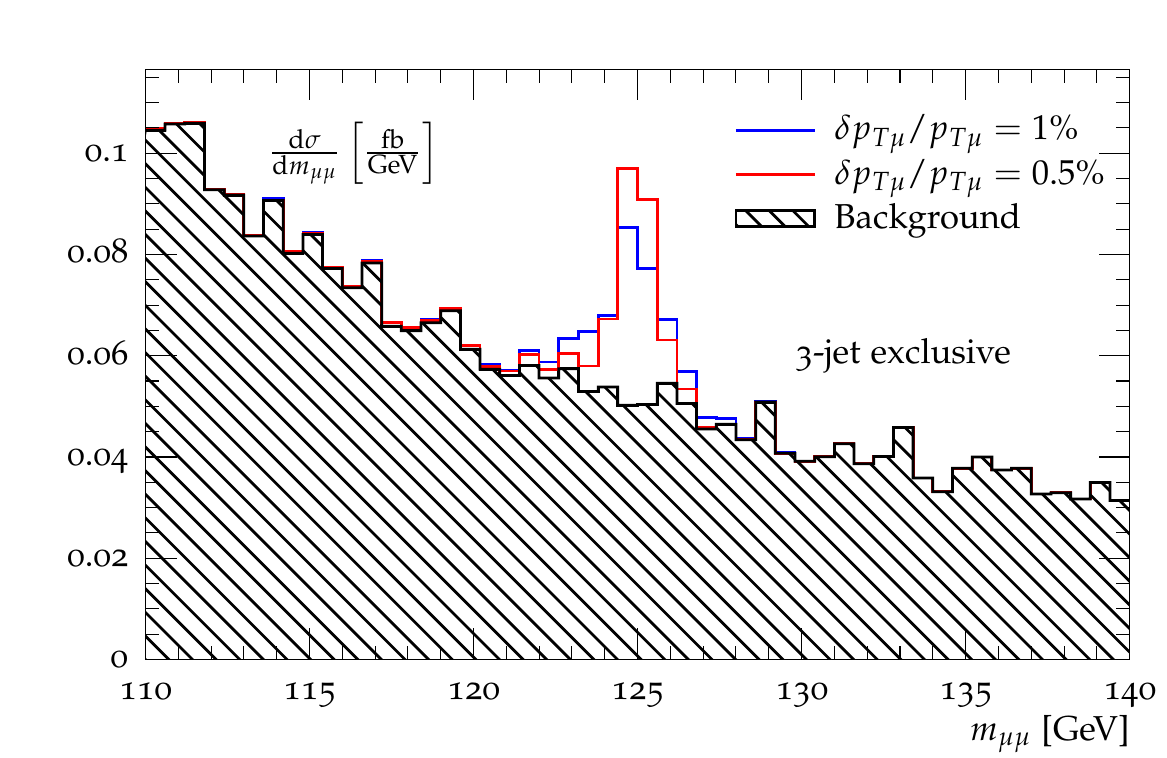}
\caption{Stacked invariant mass distribution $m_{\mu\mu}$ for the
 2-jet (left) and 3-jet (right) exclusive samples. We display
  results for two uncertainties $\delta p_{T\mu}/p_{T\mu}=0.5\%$ and
  $1\%$.}
\label{fig:mumuLineShape}
\end{figure}
%-------------------------------------------------------

%-------------------------------------------------------
\begin{figure}[t]
\includegraphics[width=0.42\textwidth]{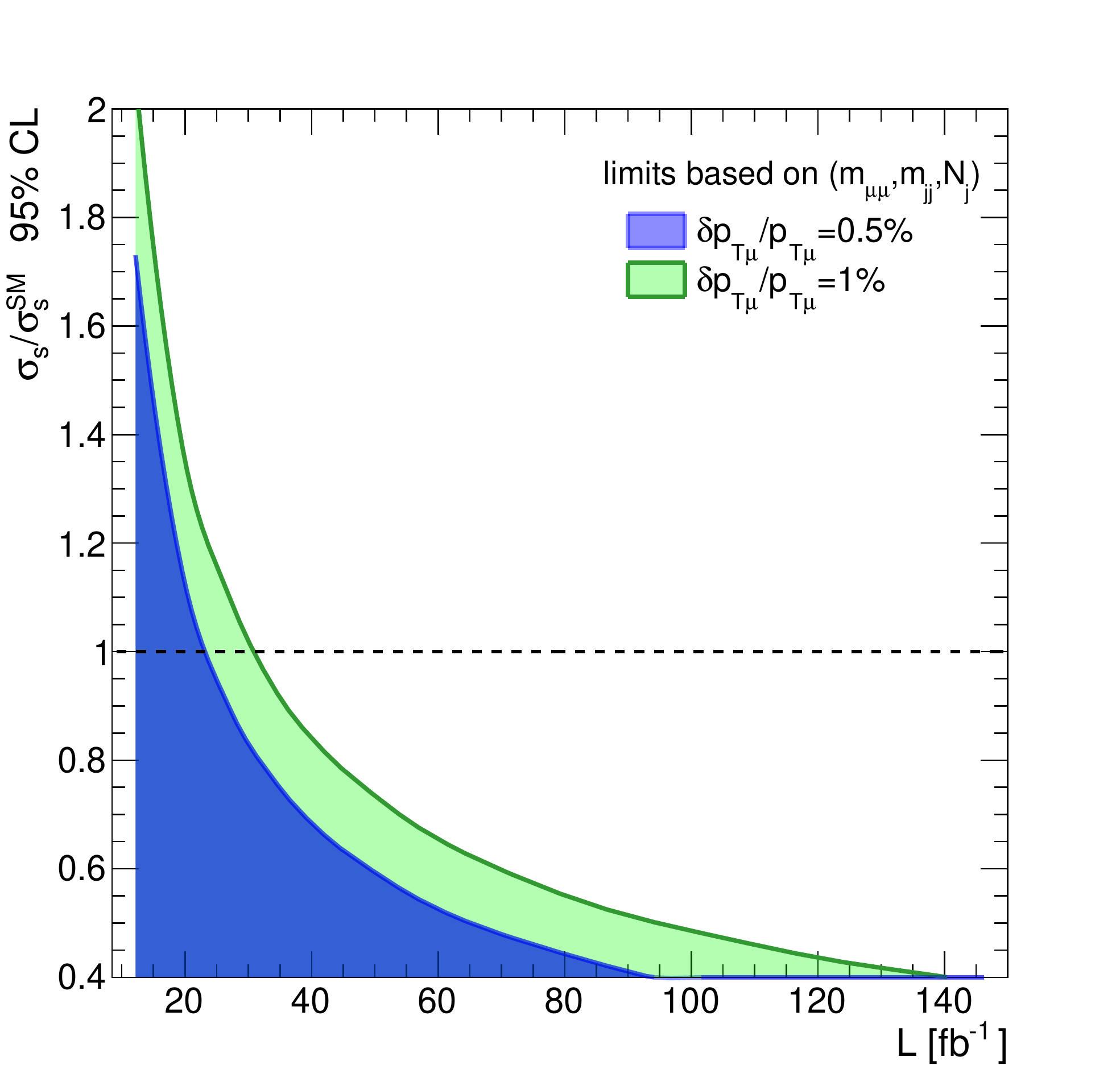}
\hspace{0.1\textwidth}
\includegraphics[width=0.42\textwidth]{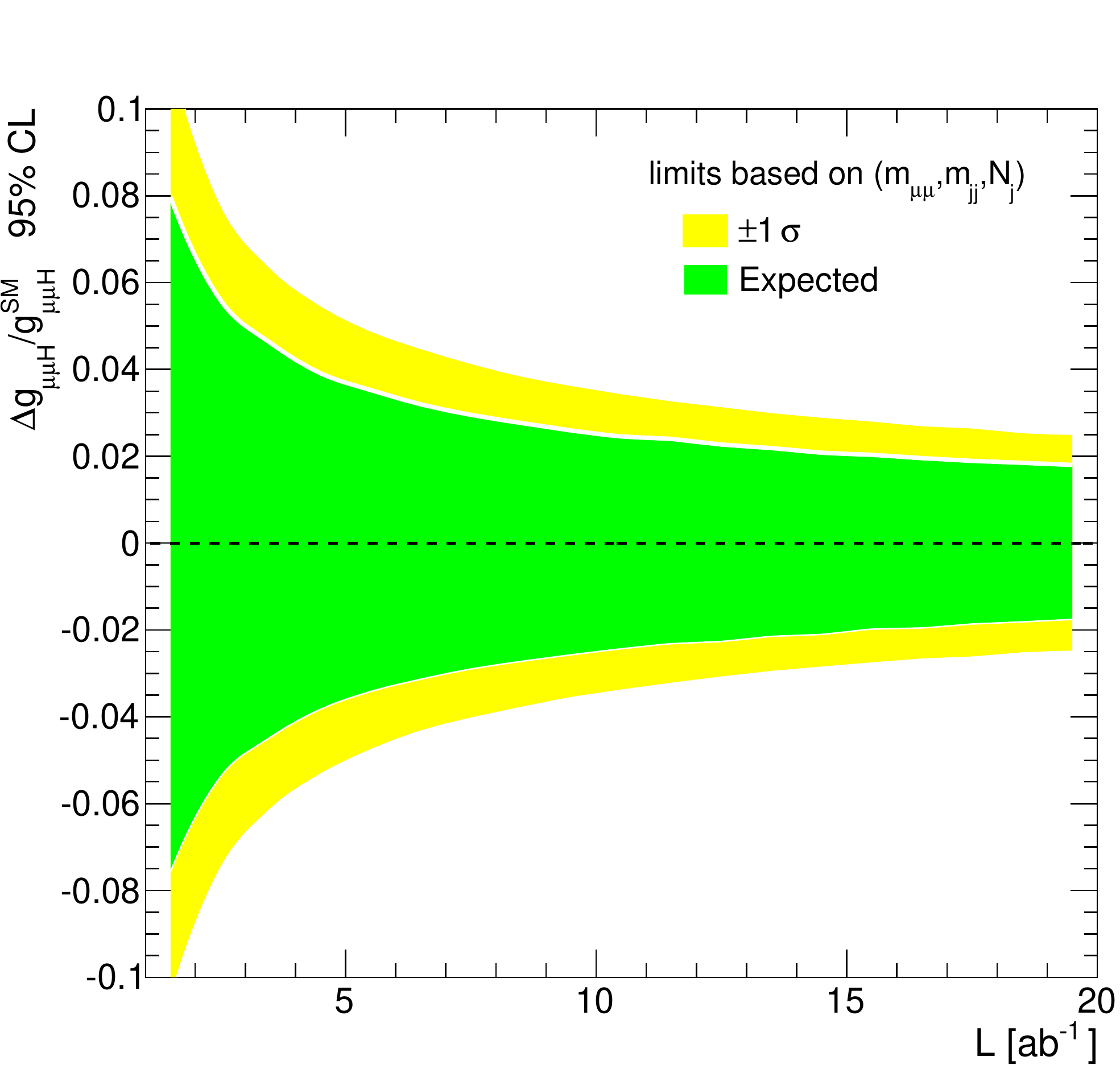}
\caption{Expected 95\% CL bound on the signal strength
  $\sigma_s/\sigma_s^\text{SM}$ (left) and on the muon coupling
  $\Delta g_{\mu \mu H}/g_{\mu \mu H}^\text{SM}$ (right) as a function
  of the integrated luminosity. We employ a log-likelihood analysis of
  the three-dimensional distribution $\{ m_{\mu\mu}, m_{j_1 j_2},
  N_j\}$. For the left panel we assume two experimental uncertainties
  for the muon reconstruction, in the right panel we fix ${\delta
    p_{T\mu}/p_{T\mu}=0.5\%}$.}
\label{fig:cls_mu}
\end{figure}
%-------------------------------------------------------

We employ a very similar analysis to the invisible Higgs search.  The
tagging jets are defined according to Eqs.\eqref{eq:ptj}
and~\eqref{eq:dphijj} and combined with the two-step jet veto defined
in Eqs.\eqref{eq:pveto1} and~\eqref{eq:eta3} to control the $t\bar{t}$
and QCD $Zjj$ backgrounds. In addition, we require a maximum amount of
missing transverse energy, $\met<40$~GeV. For the exclusive 2-jet and
3-jet samples we show the transverse momentum of the muon pair in
Fig.~\ref{fig:mjj_mu}, as one example distribution. It illustrates
how, just based on a few kinematic cuts, we will not be able to extract
the Higgs signal efficiently~\cite{lhc_muons}. Even at a 100~TeV
collider, the search for Higgs decays to muons will be a multi-variate
problem.

One of the ingredients to our analysis will still be a data-driven
side band analysis of the $m_{\mu \mu}$ distribution, searching for a
narrow Higgs peak. The stacked signal and background distribution is
shown in Fig.~\ref{fig:mumuLineShape}, for two hypothetical
experimental resolutions on the muon transverse momentum, $\delta
p_{T\mu}/p_{T\mu}=0.5\%$ and $1\%$. The muon energy scale uncertainty
directly impacts in the invariant mass resolution. At the LHC, the
typical transverse momentum uncertainty is $\delta
p_{T\mu}/p_{T\mu}\approx 1~...~2\%$ for $p_{T\mu}
=20~...~100$~GeV~\cite{muon_syst}.\medskip

As for the invisible Higgs, we again derive the sensitivity of a
100~TeV collider in the exclusive 2-jet and 3-jet bins shown in
Fig.~\ref{fig:mumuLineShape}. Similar to Eq.\eqref{eq:loglikely_inv},
we exploit the three kinematic observables
\begin{align}
\{ \, m_{\mu \mu},m_{j_1 j_2},N_j \, \} \; ,
\label{eq:loglikely_mu}
\end{align}
where the information on the Higgs resonance is encoded in $m_{\mu
  \mu}$. We show the resulting 95\%~CL sensitivity as a function of
the integrated luminosity in Fig.~\ref{fig:cls_mu}.  To quantify the
possible gains in sensitivity from a better muon momentum resolution,
we again display two hypothetical experimental uncertainties on the muon
transverse momentum, ${\delta p_{T\mu}/p_{T\mu}=0.5,1\%}$, to be
achieved at a 100~TeV collider. We conclude that it is possible to see
the $H\rightarrow \mu^+ \mu^-$ channel at 95\% CL with
$\mathcal{L}\lesssim 40~\ifb$ for both uncertainty scenarios.
Moreover, we will be able to measure the muon coupling to
\begin{align}
\frac{\Delta g_{\mu \mu H}}{g_{\mu \mu H}^\text{SM}} \lesssim 2\% \; ,
\label{eq:bottom_mu}
\end{align}
assuming a Standard Model production rate and an integrated luminosity
of $\mathcal{L}=20~\iab$.

%%%%%%%%%%%%%%%%%%%%%%%%%%%%%%%%%%%%%%%%%%%%%%%%%%
\section{Off-shell Higgs}
\label{sec:offshell}

Measuring off-shell production rates for a Higgs boson decaying to
four fermions targets a major shortcoming of any hadron collider by
giving us a handle on the total Higgs
width~\cite{offshell,offshell_ex}. The dominant LHC channel, $gg\to H^*
\to 4\ell$, is not well-suited for such a measurement, because the
gluon-Higgs coupling is loop-induced. This implies that the dependence
of the coupling has a complex dependence on the incoming and outgoing
momenta, determined by the underlying particle
content~\cite{offshell_model}. Any global analysis using off-shell
rate measurements has to be based on a well-defined
hypothesis~\cite{legacy}. If we define such a hypothesis for the
Higgs--top--gluon Lagrangian the off-shell measurements can be
naturally combined with boosted Higgs production~\cite{hjets2}.

The WBF production channel involves only renormalizable tree-level
Higgs couplings, which we know run logarithmically using the usual
renormalization group equation.  Unlike the two WBF Higgs signatures
discussed before, off-shell production in weak boson fusion is
unlikely to be seen at the LHC
altogether~\cite{offshell,offshell_model}.\medskip

%-------------------------------------------------------
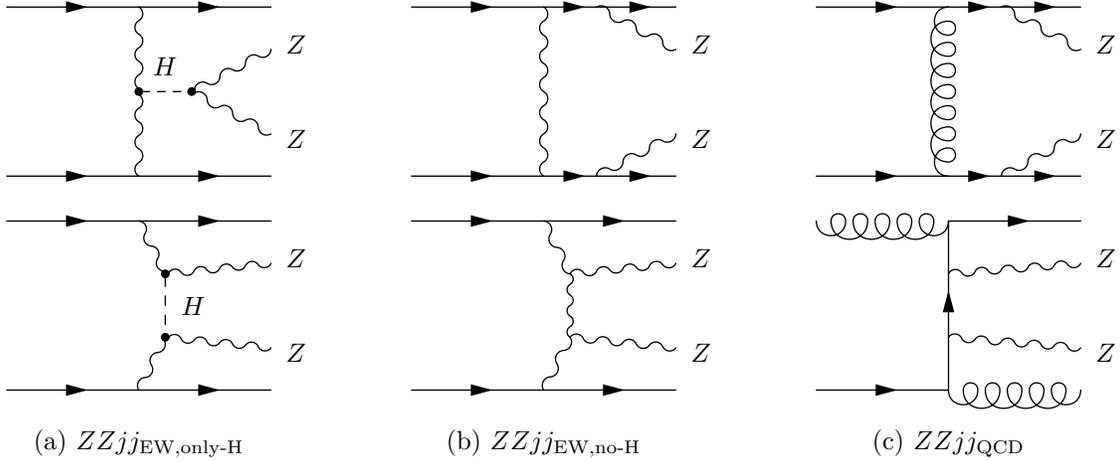
\begin{figure}[t!]
\begin{subfigure}[b]{0.24\textwidth}
%Hjj EW 1
\begin{fmfgraph*}(100,80)
\fmfset{arrow_len}{3mm}
\fmftop{q1,q3}
\fmfright{Z1,Z2}
\fmfbottom{q2,q4}
\fmflabel{{$Z$}}{Z1}
\fmflabel{{$Z$}}{Z2}
\fmf{fermion,tension=1.5,width=0.5}{q1,v1}
\fmf{fermion,width=0.5}{v1,q3}
\fmf{photon,width=0.5}{v1,v3,v4}
\fmf{dashes,tension=1.5,width=0.5,label={$H$},label.side=left}{v3,v5}
\fmfv{label.side=left,l.a=-90,l.d=.05w}{v5}
\fmfv{label.side=left,l.a=-125,l.d=.05w}{v3}
\fmfv{decor.shape=circle,decor.size=2.3}{v3}
\fmfv{decor.shape=circle,decor.size=2.3}{v5}
\fmf{fermion,tension=1.5,width=0.5}{q2,v4}
\fmf{fermion,width=0.5}{v4,q4}
\fmffreeze
\fmfforce{.5w,.9h}{v1}
\fmfforce{.5w,.1h}{v4}
\fmfforce{.7w,.5h}{v5}
\fmf{photon,tension=0.5,width=0.5}{Z1,v5,Z2}
\fmfforce{1.w,.7h}{Z1}
\fmfforce{1.w,.3h}{Z2}
\fmfforce{.5w,.5h}{v3}
\end{fmfgraph*}
%Hjj EW 2
\begin{fmfgraph*}(100,80)
\fmfset{arrow_len}{3mm}
\fmftop{q1,q3}
\fmfright{Z1,Z2}
\fmfbottom{q2,q4}
\fmflabel{{$Z$}}{Z1}
\fmflabel{{$Z$}}{Z2}
\fmf{fermion,tension=1.5,width=0.5}{q1,v1}
\fmf{fermion,width=0.5}{v1,q3}
\fmf{photon,width=0.5}{v1,v2}
\fmf{dashes,width=0.5,label={$H$},label.side=left}{v2,v3}
\fmf{photon,width=0.5}{v3,v4}
\fmf{fermion,tension=1.5,width=0.5}{q2,v4}
\fmf{fermion,width=0.5}{v4,q4}
\fmfv{label.side=left,l.a=-180,l.d=.05w}{v2}
\fmfv{label.side=left,l.a=-180,l.d=.05w}{v3}
\fmfv{decor.shape=circle,decor.size=2.3}{v2}
\fmfv{decor.shape=circle,decor.size=2.3}{v3}
\fmffreeze
\fmfforce{.5w,.9h}{v1}
\fmfforce{.5w,.1h}{v4}
\fmfforce{.6w,.35h}{v3}
\fmfforce{.6w,.65h}{v2}
\fmfforce{1.w,.7h}{Z1}
\fmfforce{1.w,.3h}{Z2}
\fmf{photon,width=0.5}{v3,Z2}
\fmf{photon,width=0.5}{v2,Z1}
\end{fmfgraph*}
\caption{$ZZjj_\text{EW,only-H}$}
\end{subfigure}
\hspace{1.2cm}
 %Zjj EW
 %diagram3
\begin{subfigure}[b]{0.24\textwidth}
\begin{fmfgraph*}(100,80)
\fmfset{arrow_len}{3mm}
\fmftop{q1,q3}
\fmfright{Z1,Z2}
\fmfbottom{q2,q4}
\fmflabel{{$Z$}}{Z1}
\fmflabel{{$Z$}}{Z2}
\fmf{fermion,tension=1.5,width=0.5}{q1,v1,v5,q3}
\fmf{photon,width=0.5}{v1,v4}
\fmf{fermion,tension=1.5,width=0.5}{q2,v4,v3,q4}
\fmffreeze
\fmfforce{.5w,.9h}{v1}
\fmfforce{.7w,.9h}{v5}
\fmfforce{.7w,.1h}{v3}
\fmfforce{.5w,.1h}{v4}
\fmf{photon,tension=0.5,width=0.5}{v5,Z2}
\fmf{photon,tension=0.5,width=0.5}{v3,Z1}
\fmfforce{1.w,.7h}{Z2}
\fmfforce{1.w,.3h}{Z1}
\end{fmfgraph*}
%diagram4
\begin{fmfgraph*}(100,80)
\fmfset{arrow_len}{3mm}
\fmftop{q1,q3}
\fmfright{Z1,Z2}
\fmfbottom{q2,q4}
\fmflabel{{$Z$}}{Z1}
\fmflabel{{$Z$}}{Z2}
\fmf{fermion,tension=1.5,width=0.5}{q1,v1}
\fmf{fermion,width=0.5}{v1,q3}
\fmf{photon,width=0.5}{v1,v2}
\fmf{photon,width=0.5}{v2,v3}
\fmf{photon,width=0.5}{v3,v4}
\fmf{fermion,tension=1.5,width=0.5}{q2,v4}
\fmf{fermion,width=0.5}{v4,q4}
\fmffreeze
\fmfforce{.5w,.9h}{v1}
\fmfforce{.5w,.1h}{v4}
\fmfforce{.6w,.35h}{v3}
\fmfforce{.6w,.65h}{v2}
\fmfforce{1.w,.7h}{Z1}
\fmfforce{1.w,.3h}{Z2}
\fmf{photon,width=0.5}{v3,Z2}
\fmf{photon,width=0.5}{v2,Z1}
\end{fmfgraph*}
\caption{$ZZjj_\text{EW,no-H}$}
\end{subfigure}
\hspace{1.2cm}
%%Zjj QCD
 %diagram5
\begin{subfigure}[b]{0.24\textwidth}
\begin{fmfgraph*}(100,80)
\fmfset{arrow_len}{3mm}
\fmftop{q1,q3}
\fmfright{Z1,Z2}
\fmfbottom{q2,q4}
\fmflabel{{$Z$}}{Z1}
\fmflabel{{$Z$}}{Z2}
\fmf{fermion,tension=1.5,width=0.5}{q1,v1,v5,q3}
\fmf{gluon,width=0.5}{v1,v4}
\fmf{fermion,tension=1.5,width=0.5}{q2,v4,v3,q4}
\fmffreeze
\fmfforce{.5w,.9h}{v1}
\fmfforce{.7w,.9h}{v5}
\fmfforce{.7w,.1h}{v3}
\fmfforce{.5w,.1h}{v4}
\fmf{photon,tension=0.5,width=0.5}{v5,Z2}
\fmf{photon,tension=0.5,width=0.5}{v3,Z1}
\fmfforce{1.w,.7h}{Z2}
\fmfforce{1.w,.3h}{Z1}
\end{fmfgraph*}
%diagram6
\begin{fmfgraph*}(100,80)
\fmfset{arrow_len}{3mm}
\fmftop{g1,q3}
\fmfright{Z1,Z2}
\fmfbottom{q2,g4}
\fmflabel{{$Z$}}{Z1}
\fmflabel{{$Z$}}{Z2}
\fmf{gluon,tension=1.5,width=0.5}{g1,v1}
\fmf{fermion,tension=1.5,width=0.5}{q2,v4}
\fmf{plain,width=0.5}{v4,v3}
\fmf{fermion,width=0.5}{v3,v2}
\fmf{plain,width=0.5}{v2,v1}
\fmf{fermion,width=0.5}{v1,q3}
\fmf{gluon,width=0.5}{v4,g4}
\fmffreeze
\fmfforce{.5w,.9h}{v1}
\fmfforce{.5w,.1h}{v4}
\fmfforce{.5w,.35h}{v3}
\fmfforce{.5w,.65h}{v2}
\fmfforce{1.w,.7h}{Z1}
\fmfforce{1.w,.3h}{Z2}
\fmf{photon,width=0.5}{v3,Z2}
\fmf{photon,width=0.5}{v2,Z1}
\end{fmfgraph*}
\caption{$ZZjj_\text{QCD}$}
\end{subfigure}
 \caption{Representative set of Feynman diagrams for WBF $ZZjj$
   production through a Higgs (left), in an electroweak process
   (center), and including the strong interaction (right).}
 \label{fig:feyn_offshell}
\end{figure}
%-------------------------------------------------------

At a 100~TeV collider we expect a sizeable event sample for $pp\to H
\, jj_\text{EW} \to (4\ell) \, jj$, even for off-shell
Higgs production with $m_{4 \ell} \gg m_H$.  The dominant backgrounds
are $ZZjj_\text{QCD}$ and $ZZjj_\text{EW}$ production, illustrated in
Fig.~\ref{fig:feyn_offshell}. There is an interference between the
EW and QCD amplitudes, but it is color suppressed and has been shown
to be negligible for total rates and for distributions after the WBF
cuts~\cite{WBF_QCD_EW}. As always, the QCD background can be
suppressed by the standard WBF cuts and a central jet veto. Unlike for
on-shell Higgs decays, the EW background helps our analysis
through its interference with the Higgs diagrams.  We generate the
signal and background samples with
\textsc{MadGraph5+Pythia8}~\cite{mg5,pythia8} and observe good
agreement with \textsc{MCFM}~\cite{Campbell:2015vwa}. Spin
correlations and off-shell effects are fully accounted for, including
the $Z$-decays.

We start by requiring four isolated leptons with
less than 15\% of the hadronic activity within a radius of $R=0.2$.
The kinematic selection follows the CMS analysis~\cite{offshell_ex},
\begin{alignat}{9}
p_{T\ell}&>5~\gev \qqqquad   & |\eta_\ell|&<2.5 \qqqquad & m_{4\ell}&>100~\gev \notag \\
m_{\ell \ell'}&>4~\gev \qqqquad
& m_{\ell \ell,1} &= [40,120]~\gev \qqqquad 
& m_{\ell \ell,2} &= [12,120]~\gev  \; .
\label{eq:m4l} 
\end{alignat}
The two $m_{\ell \ell}$ ranges define a leading and a sub-leading
flavor-matched lepton pair. For the tagging jets we again require
Eq.\eqref{eq:ptj}, combined with $m_{j_1 j_2}>600$~GeV and our
two-step jet veto based on Eqs.\eqref{eq:pveto1}
and~\eqref{eq:eta3}.\medskip

%-------------------------------------------------------
\begin{figure}[!t]
\centering
 \includegraphics[width=0.35\textwidth]{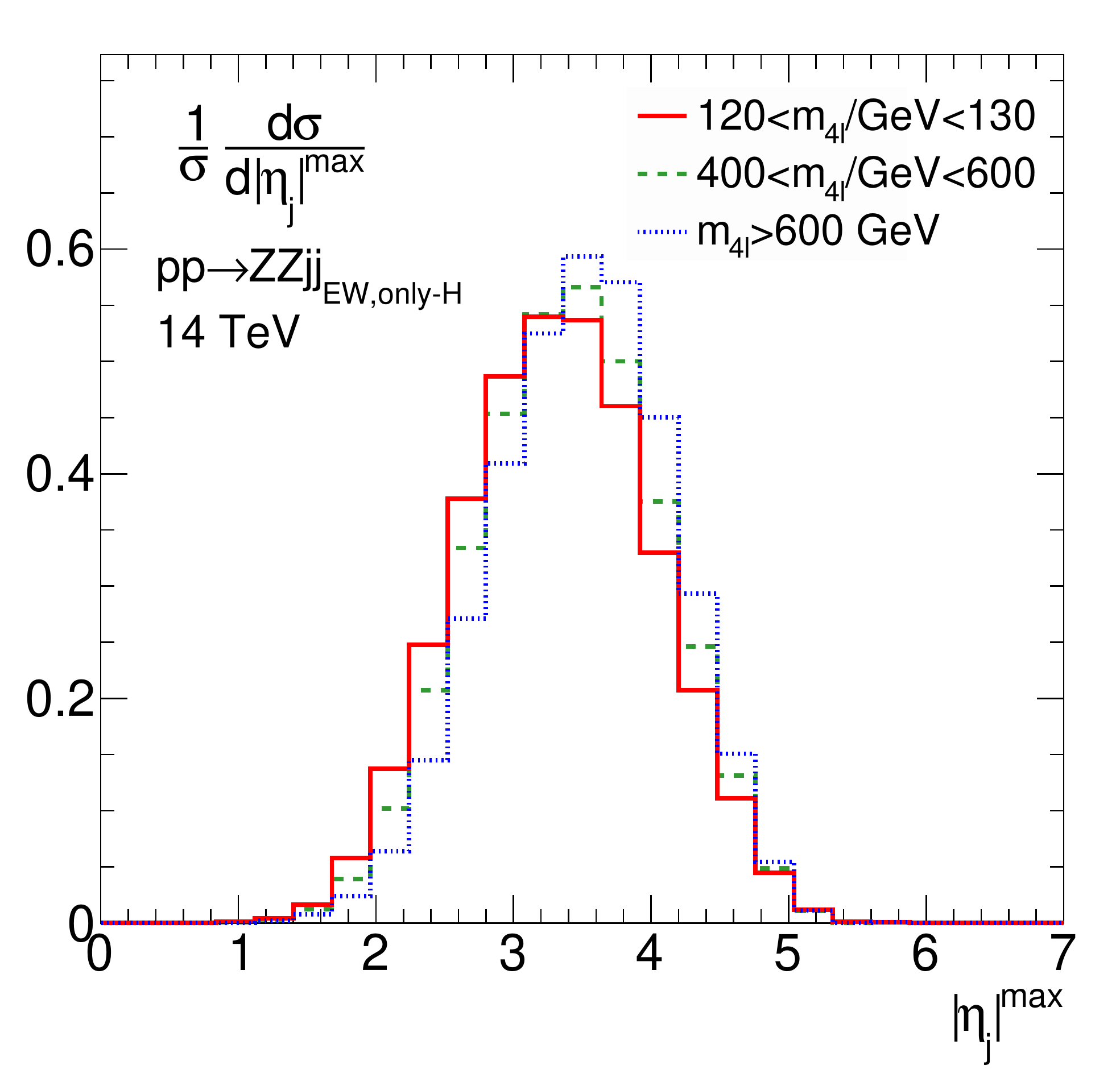}
 \hspace{0.1\textwidth}
 \includegraphics[width=0.35\textwidth]{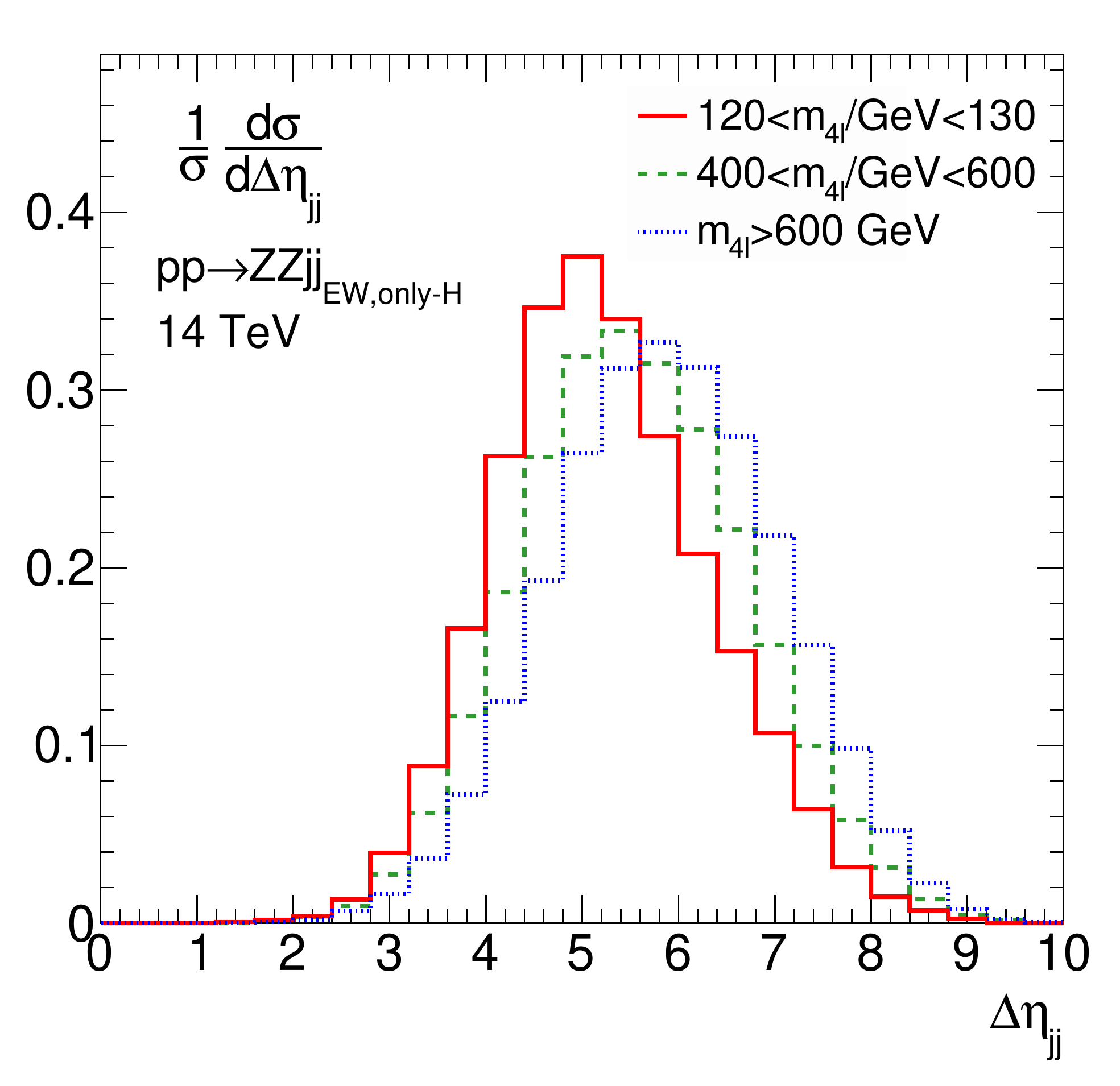}
 \includegraphics[width=0.35\textwidth]{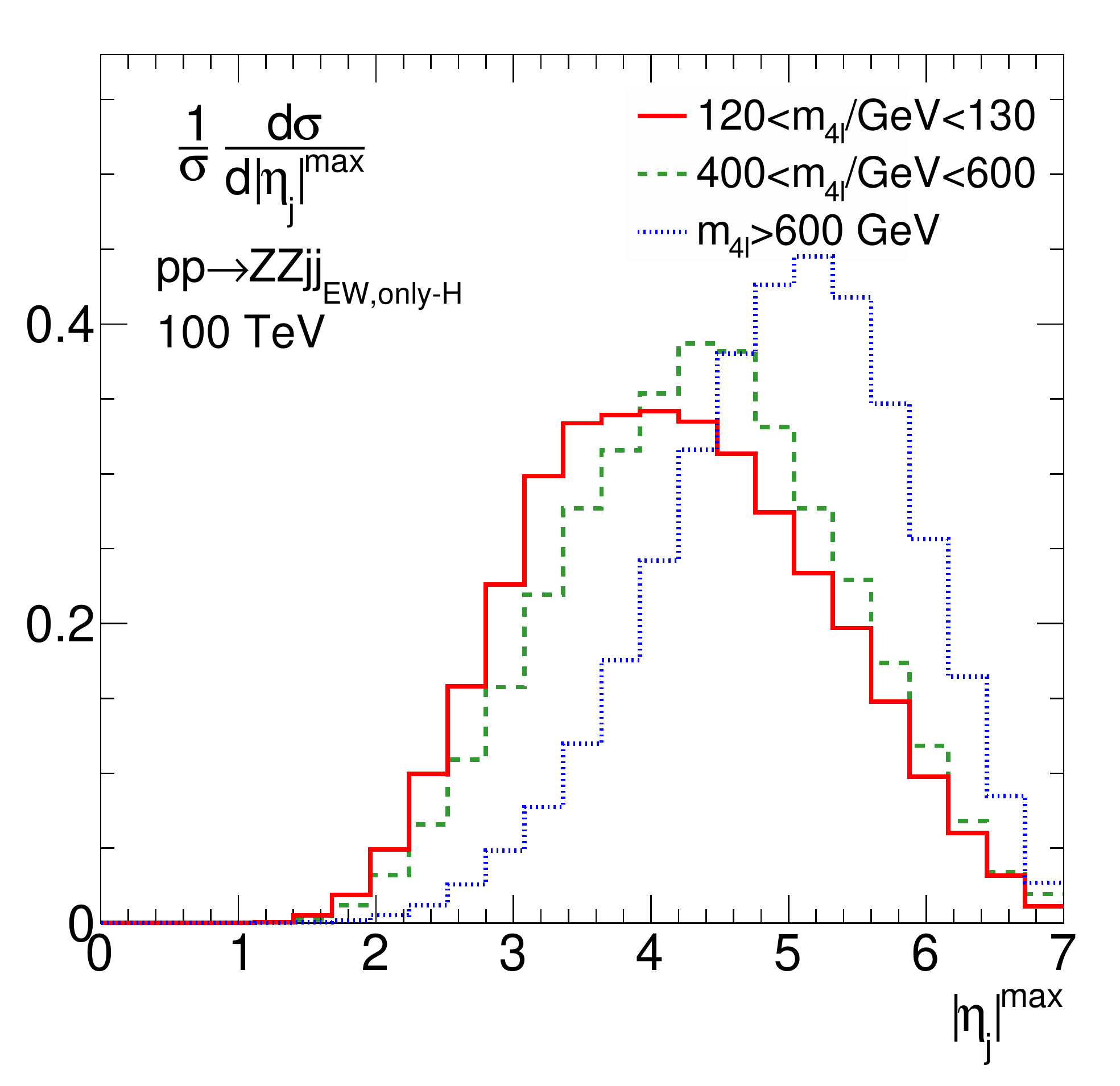}
 \hspace{0.1\textwidth}
 \includegraphics[width=0.35\textwidth]{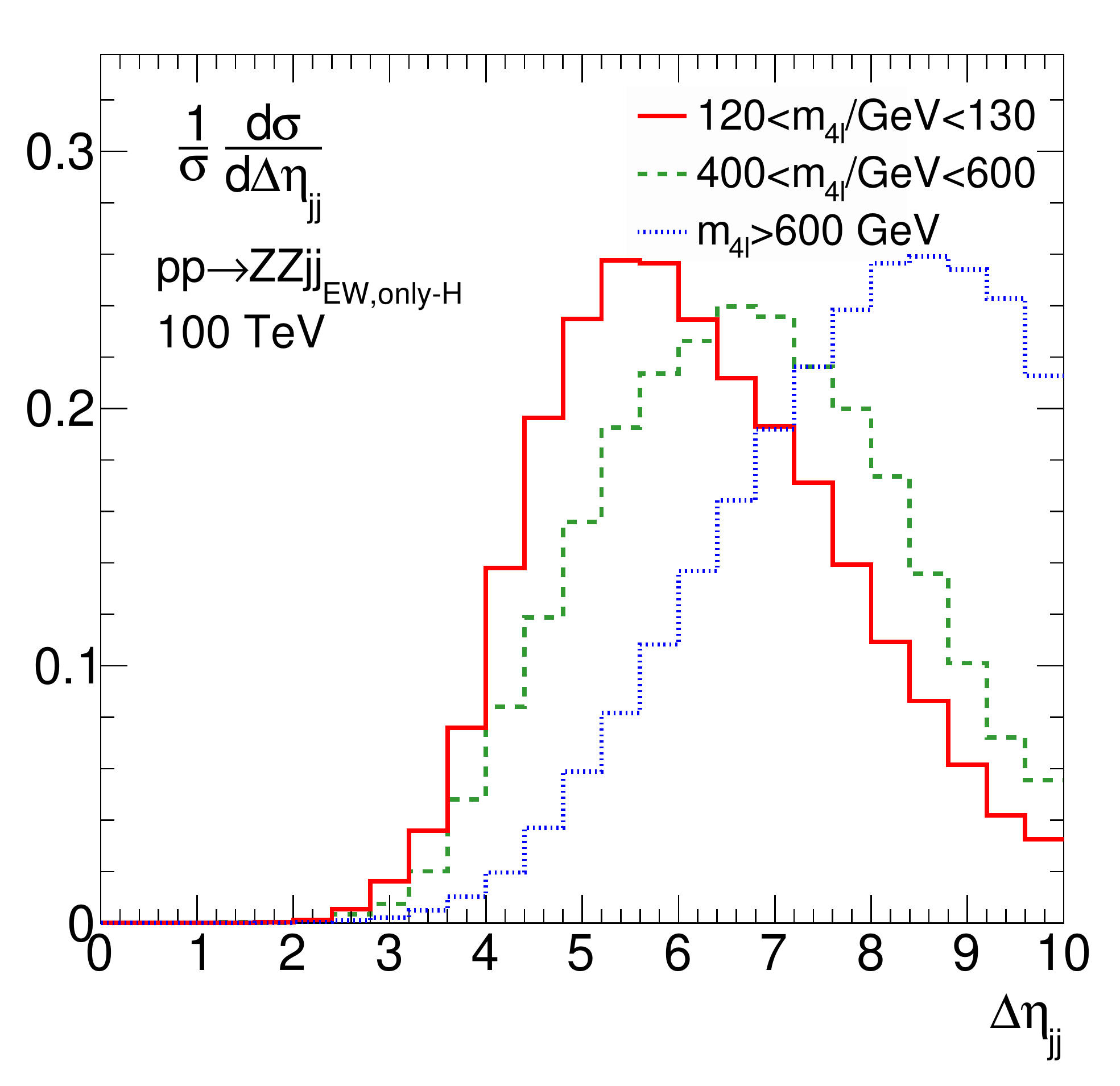}
\caption{Leading tagging jet rapidity (left) and rapidity difference
  (right) for $ZZjj_\text{EW,only-H}$ signal at 14~TeV (top) and
  100~TeV (bottom).  We show three phase space regions not
  requiring the standard cuts $|\eta_{j_{1,2}}|<5$ and $\Delta \eta_{j_1 j_2}>5$.}
\label{fig:etajj_off}
\end{figure}
%-------------------------------------------------------

Building on the tagging jet kinematics discussed in
Sec.~\ref{sec:jets}, we show the tagging jet rapidities for three
slices of $m_{4\ell}$ in Fig.~\ref{fig:etajj_off}.  For more off-shell
Higgs production, the tagging jets move further into the forward region.
This behavior is related to gauge boson scattering, $VV\rightarrow VV$,
at high energies.  The off-shell phase space region provides the ideal
setup for the effective $W/Z$-approximation, where the vector boson
parton picture requires a hierarchy of energy scale, ${\sqrt{s}\gg
  m_{4\ell}\gg m_V}$.  Here, the longitudinal and transverse
scattering amplitudes scale as $\mathcal{A}_{LL}/\mathcal{A}_{TT} \sim
m_{4\ell}^2/m_V^2$.  This feature is more prominent at 100~TeV than at
14~TeV collider energy because at larger scattering energies we
produce a greater fraction of longitudinal gauge boson even at the
Higgs pole~\cite{splitting2}, see Fig.~\ref{fig:eta}.  We can use this
feature by increasing the $\Delta\eta_{j_1 j_2}$ cut for the off-shell
analysis, indicating that a detector coverage to rapidities larger
than $|\eta_j|=5$ will be even more important than in other WBF
channels.\medskip

In what follows, we will assume that the Higgs couplings to $W$ and
$Z$ gauge bosons change simultaneously as
$g_{ZZH}/g_{ZZH}^\text{SM}=g_{WWH}/g_{WWH}^\text{SM}$, respecting
custodial symmetry. We can then write the $ZZjj_\text{EW}$ amplitude as the
sum of the $ZZjj_\text{EW,only-H}$ and $ZZjj_\text{EW,no-H}$
contributions,
\begin{alignat}{5}
\mathcal{A}_\text{EW}= 
\left( \frac{g_{ZZH}}{g_{ZZH}^\text{SM}} \right)^2 
 \mathcal{A}_H+\mathcal{A}_B \;,
\label{eq:amp_offshell} 
\end{alignat}
where $\mathcal{A}_H$ corresponds to Fig.~\ref{fig:feyn_offshell}(a)
and $\mathcal{A}_B$ corresponds to
Fig.~\ref{fig:feyn_offshell}(b). Following this notation, any
observable distribution can be decomposed as
\begin{alignat}{5}
\frac{d\sigma_\text{EW}}{d\mathcal{O}}=
 \left( \frac{g_{ZZH}}{g_{ZZH}^\text{SM}} \right)^4
  \frac{d\sigma_{HH}}{d\mathcal{O}}+
 \left( \frac{g_{ZZH}}{g_{ZZH}^\text{SM}} \right)^2
  \frac{d\sigma_{HB}}{d\mathcal{O}}+ 
                          \frac{d\sigma_{BB}}{d\mathcal{O}} \; .
\label{eq:sigma_offshell} 
\end{alignat}
While any measurement of an on-shell Higgs signal rate has a flat
direction if we vary the involved Higgs couplings together with the
total Higgs width, the above measurement will allow us to break this
degeneracy and derive a bound on $\Gamma_H$.\medskip

%-------------------------------------------------------
\begin{figure}[t!]
\includegraphics[width=0.38\textwidth]{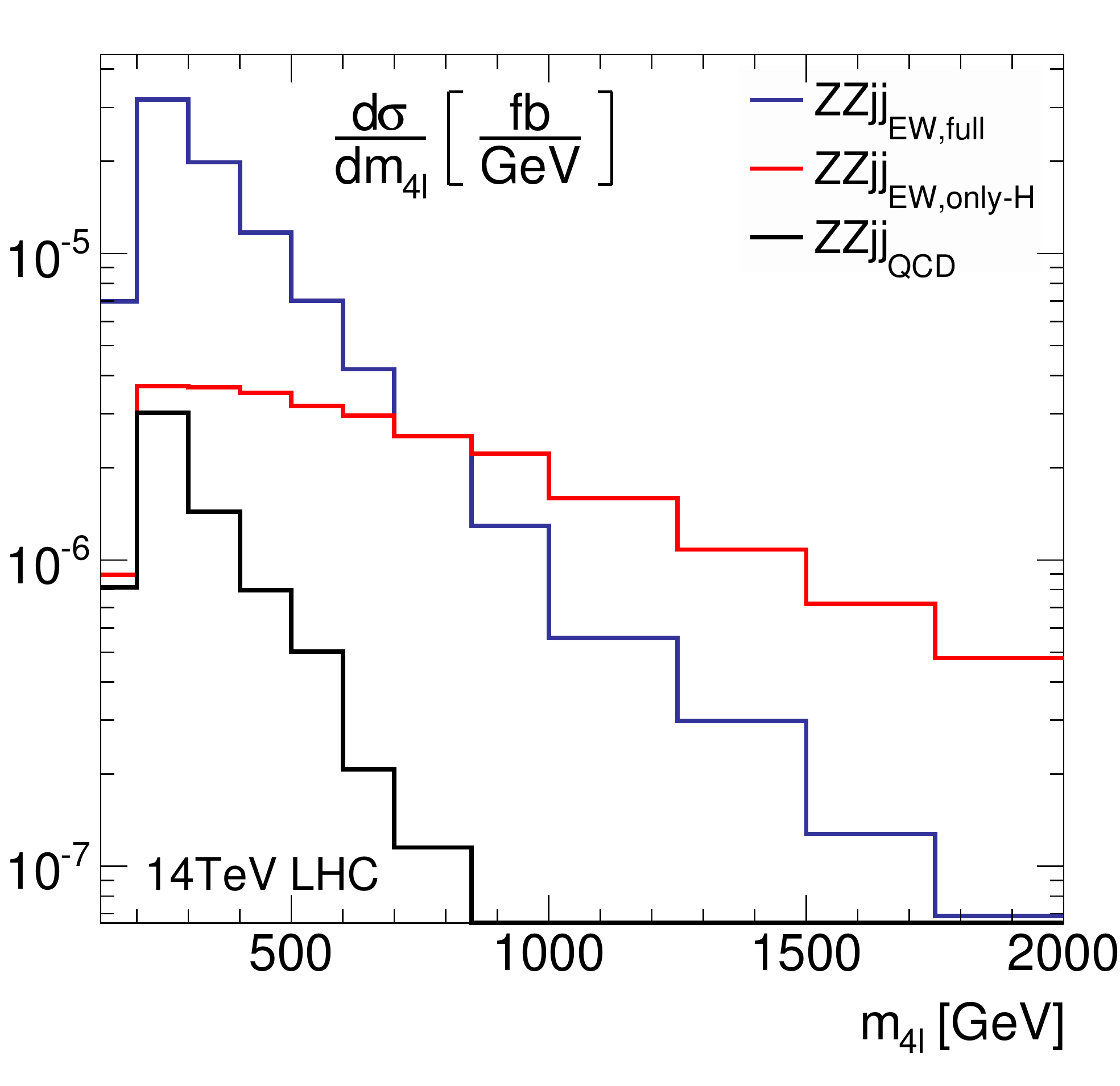}
 \hspace{0.1\textwidth}
\includegraphics[width=0.38\textwidth]{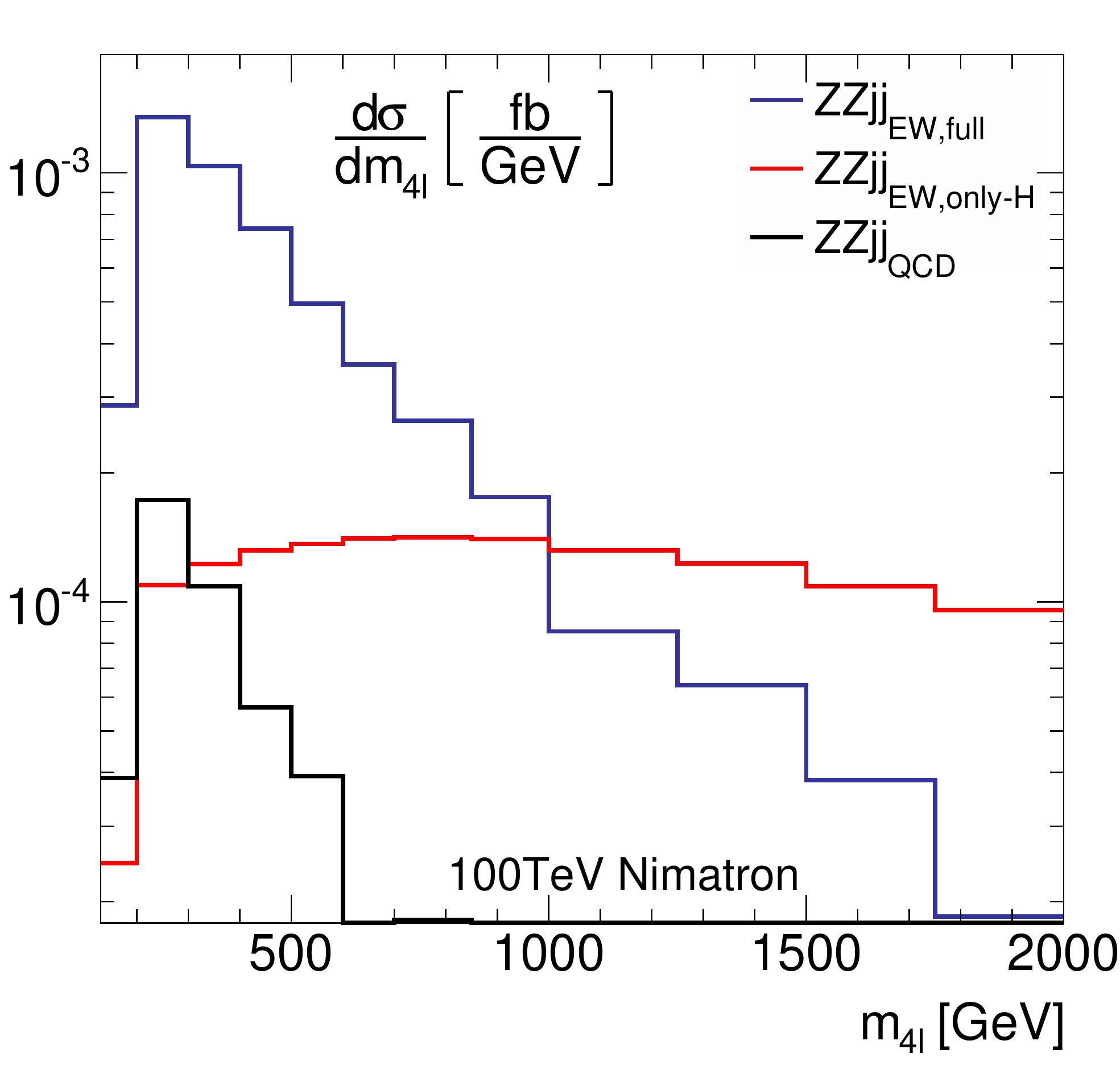}
\caption{$m_{4\ell}$ distributions for the Higgs signal and the
  different backgrounds at 14~TeV (left) and 100~TeV (right).  We
  require $\Delta \eta_{j_1 j_2}>6$ to enhance the signature.}
% Instead of this crude cut on $\Delta \eta_{j_1 j_2}$, our final bounds will 
%    explore the larger rapidity separation kinematics via a binned log-likelihood analysis of $(m_{4l},\Delta \eta_{j_1 j_2})$.}
\label{fig:m4l}
\end{figure}
%-------------------------------------------------------

%-------------------------------------------------------
\begin{figure}[!b]
  \includegraphics[width=.43\textwidth]{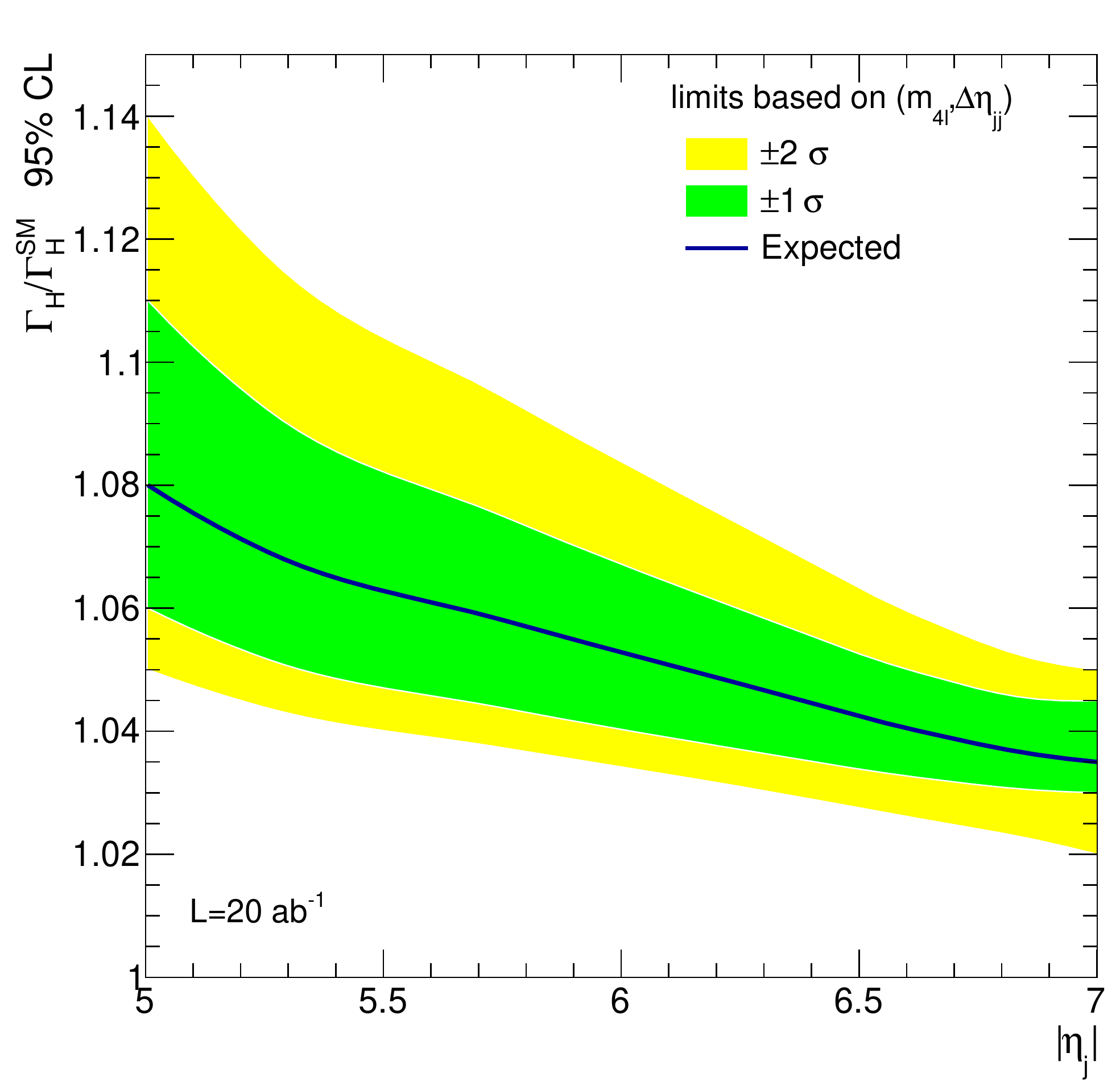}
\caption{Expected 95\% bound on the total Higgs width, based on a
  log-likelihood analysis of the two-dimensional distribution
  $\{ m_{4\ell},\Delta\eta_{j_1 j_2}  \}$.}
\label{fig:cls_m4l}
\end{figure}
%-------------------------------------------------------

In Fig.~\ref{fig:m4l} we show the $m_{4\ell}$ distribution for the
signal and backgrounds at 14~TeV (left) and 100~TeV (right). First, we
observe that the QCD background, $ZZjj_\text{QCD}$, is depleted by our
selections, leaving $ZZjj_\text{EW}$ as the leading background.
Secondly, the interference between the $ZZjj_\text{EW,only-H}$ signal
and the background $ZZjj_\text{EW,no-H}$ is large and destructive.
This leads to a smaller full $ZZjj_\text{EW,full}$ rate than the Higgs
signal alone in the far off-shell regime. Finally, the
signal distribution at 100~TeV presents a  significantly smaller slope than at
the LHC, related to the stronger longitudinal gauge boson polarization
at larger energies.\medskip

Again, we derive our bound on the measured total Higgs with in terms
of $\Gamma_H/\Gamma_H^\text{SM}$ through a log-likelihood analysis. While
the Higgs width can be either smaller or larger than the SM
prediction, the more interesting question is how we can constrain
additional, unobserved Higgs decay channels, leading to an increase in
the width, $\Gamma_H/\Gamma_H^\text{SM} >1$. Because the signal and the
leading background are both electroweak, we do not need to include
$N_j$ as part of this analysis. In addition, following the above
arguments we replace $m_{j_1 j_2}$ from the previous analyses in
Eq.\eqref{eq:loglikely_inv} and \eqref{eq:loglikely_mu} by $\Delta
\eta_{j_1 j_2}$, giving us a likelihood distribution over 
\begin{align}
\{ \, m_{4\ell},\Delta\eta_{j_1 j_2} \, \} \; .
\label{eq:loglikely_off}
\end{align}
For consistency, we discard events with ${m_{4\ell}>2}$~TeV to avoid
gathering sensitivity from unitarity-violating theory
predictions~\cite{Lee:1977eg}. Our projected reach and the associated
uncertainties are displayed as a function of the rapidity coverage in
Fig.~\ref{fig:cls_m4l}.  We find that a 100~TeV collider will be
sensitive to
\begin{align} 
\frac{\Gamma_H}{\Gamma_H^\text{SM}} > 
\begin{cases}
1.08 \qquad & \text{for} \; |\eta_j|<5 \notag \\ 
1.04 \qquad & \text{for} \; |\eta_j|<6.5 \; ,
\end{cases} 
\end{align}
assuming a Standard Model production rate and an integrated luminosity
of $\mathcal{L}=20~\iab$.

%%%%%%%%%%%%%%%%%%%%%%%%%%%%%%%%%%%%%%%%%%%%%%%%%%
\section{Summary}
\label{sec:summary}

We have systematically studied Higgs production in weak boson fusion
at a future 100~TeV hadron collider. This signature is 
crucial for global analyses of the Higgs sector and gives us
access to non-trivial Higgs properties. We started with
an analysis of the tagging jet kinematics, indicating that it would be
beneficial to extend the calorimeter coverage to rapidities $\eta_j
\approx 6$. A central jet analysis significantly reduced all
QCD backgrounds and ensured that the dominant Higgs signal at a
100~TeV hadron collider is from weak boson fusion. We advertize a
two-step veto, where a third jet in between the two
tagging jets is kinematically analyzed, while a fourth jet is
vetoed. The details of the tagging jet kinematics and the rate
dependence on the jet veto scale will allow us to reduce the
dependence of WBF rate measurements from Monte Carlo
simulations.\medskip

Instead of a global Higgs couplings analysis, where most
underlying rate measurements at 100~TeV will be systematics or theory
limited and the quantitative results will be just guess work, we
studied three particularly challenging WBF benchmark
signatures.

First, we studied Higgs decays to invisible states, for
example dark matter candidates in Higgs portal models. We found that,
depending on experimental systematics, we can test invisible Higgs
decays with a branching ratio from one per-cent to one per-mille at a
100~TeV collider with an integrated luminosity of $20~\iab$. The key
challenge in this analysis, as well as in the corresponding HL-LHC
analysis, is our understanding of the central hadronic
activity~\cite{cathy}.

Next, we determined the reach of a 100~TeV collider in measuring the
muon Yukawa coupling. This analysis rests on our ability to separate
the kinematically similar $Z$-decays to muon pairs from the Higgs
signal. A precision measurement of the muon Yukawa coupling, assuming
a Standard Model production cross section, should be at the level of
2\%.

Finally, we considered a measurement of the total Higgs width through
off-shell Higgs production in weak boson fusion. Unlike gluon fusion
production, this signature does not have a significant model
dependence, because the underlying coupling appears at tree level and
is therefore renormalizable. We find that a 100~TeV will be able to
detect an enhancement of the total Higgs width by around 5\%.\medskip

All of these Higgs precision analyses will hugely benefit from
dedicated WBF triggers and an increased tagging jet rapidity coverage
at a 100~TeV collider. Under realistic assumptions the WBF processes
should allow us to systematically study electroweak processes at an
energy-frontier hadron collider.

\bigskip
\begin{center} \textbf{Acknowledgments}
\end{center}

DG was funded by U.S.  National Science Foundation under the grant
PHY-1519175.  JT is supported by the \textsl{Bundesministerium für
  Bildung und Forschung} (BMBF Grant No.~05H15VHCA1).

%%%%%%%%%%%%%%%%%%%%%%%%%%%%%%%%%%%%%%%%%%%%%%%%

\end{fmffile}
\end{document}